\documentclass[showpacs,preprint,amsmath,PRB]{revtex4}
\usepackage{graphicx}
\usepackage{dcolumn}
\usepackage{bm}

\begin{document}

\title{On the quantum master equation for fermions }
\author{ Chun-Feng Huang${^{(1),(2)}}$ and Keh-Ning Huang${^{(1),(3)}}$}

\affiliation{$^{(1)}$ Department of Physics, National Taiwan University, Taipei, Taiwan, R. O. C.}
\affiliation{$^{(2)}$ National Measurement Laboratory, Center for Measurement Standards, Industrial Technology Research Institute, Hsinchu, Taiwan, R. O. C. }
\affiliation{$^{(3)}$ Institute of Atomic and Molecular Sciences, Academia Sinica, Taipei, Taiwan, R. O. C.}

\date{\today}

\begin{abstract}
A quantum master equation is obtained for identical fermions by including a relaxation term in addition to the mean-field Hamiltonian. [Huang C F and Huang K N 2004 Chinese J. Phys. ${\bf 42}$ 221; Gebauer R and Car R 2004 Phys. Rev. B ${\bf 70}$ 125324] It is proven in this paper that both the positivity and Pauli's exclusion principle are preserved under this equation when there exists an upper bound for the transition rate. Such an equation can be generalized to model BCS-type quasiparticles, and is reduced to a Markoff master equation of Lindblad form in the low-density limit with respect to particles or holes.   
\end{abstract}

\maketitle

\section{Introduction}  

The quantum kinetic approaches have attracted much attention because of the development of the nanoscale devices, for which the classical kinetic theory may become invalid. \cite{Shah,other1,other2,Callebaut,Ohtsuki,Burke,Iotti} Different master equations have been introduced to develope the quantum kinetic models. \cite{Ohtsuki,Burke,Iotti,Huang,Cui,Ralph,Hubner,Geoke,Blum,Louisell,Kampen,Lindblad,Giulini} In these equations, there are relaxation terms responsible for the irresversibility. The Markoff master equation \cite{Ohtsuki,Louisell} is used to describe quantum relaxation processes, and the master equation of Lindbald form \cite{Kampen,Lindblad} is derived based on mathematical assumptions. After ignoring the pure-dephasing term \cite{Louisell}, the Markoff master equation is just a particular master equation of Lindblad form. The well-known Pauli master equation, in fact, can be deduced from the Markoff master equation in the incoherent limit. \cite{Louisell}

It is important to assume the linearity to derive the master equation of Lindblad form in quantum mechanics. \cite{Kampen} On the other hand, nonlinear relaxation terms have been introduced semiclassically based on Pauli's exclusion principle for systems composed of identical fermions. \cite{Liboff,Singh} A nonlinear quantum master equation has been introduced in Refs. \cite{Huang} and \cite{Ralph} to unify the quantum and semiclassical approaches. The relaxation term of such an equation, in fact, can be constructed by considering two antihermitian terms. \cite{Huang} One is responsible for the loss of particles while the other is for the loss of holes, which is equivalent to the gain of particles. Here holes are vacencies of any orbitals. \cite{Huang,Singh} Therefore, the relaxation term is symmetric with respect to particles and holes. After incorporating a pairing tensor, the density matrices for particles and holes can be used to constructed those for quasiparticles in Bardeen-Cooper-Schrieffer (BCS) pairing models \cite{Paar,Valatin}, which include many-body effects beyond one-body approximation. Different approaches \cite{Laughlin_FQHE,MCRRPA,exciton} have been developed to study many-body correlations, and BCS pairing models are powerful to understand superconductivity \cite{superconductivity}, superfluid \cite{superconductivity2}, and meson-nucleon couplings \cite{Paar}.     

It will be proven in this paper that the nonlinear master equation introduced by Refs. \cite{Huang} and \cite{Ralph} preserves both the positivity and Pauli's exclusion principle when there exists an upper bound for the transition rate. Therefore, such an equation is suitable to model fermions. For convenience, first we discuss different types of master equations in section II, and the proof is in section III. As discussed in section IV, the nonlinear master equation for fermions can be reduced to the Markoff master equation in the low-density limit with respect to particles or holes. It is discussed how to prove the conservation of the trace of the density matrix. An extension of the master equation to the relativisitic Hartree-Bogoliubov model \cite{Paar,Valatin}, which is a BCS-pairing model, is obtained by introducing a constraint on the relaxation term. It is also mentioned in section IV how to consider the re-pairing between BCS-type quasiparticles and quasiholes to incorporate multiple order parameters \cite{superconductivity2,heavy_fermion}. We note that multiple order parameters have been introduced not only to understand heavy fermion metals \cite{heavy_fermion}, but also to unify BCS theory and ferromagnetic/antiferromagnetic theory to probe high-temperature superconductors. \cite{Laughlin} Conclusions are made in section V.

\section{Different types of master equations}

Different types of master equations have been introduced to describe irreversible processes. \cite{Ohtsuki,Burke,Iotti,Huang,Cui,Ralph,Hubner,Geoke,Blum,Louisell,Kampen,Lindblad,Giulini,Liboff,Singh,Entin} For a quantum system with the density matrix $\rho (t)$, the master equation of the following form 
\begin{eqnarray}
\frac{\partial }{\partial t}\rho (t)=i[\rho (t),H(t)]+R(\rho (t)) 
\end{eqnarray}
has been discussed in the literature. Here $R(\rho (t))$ is the relaxation term, $[A,B]$ denotes the commutator of any two operators $A$ and $B$, and $H(t)$ is the Hamiltonian generating a unitary operator $U(t)$ by 
\begin{eqnarray}
i\frac{\partial}{\partial t} U(t) = H (t) U(t)  
\end{eqnarray}
with $U(t _{0})= I$, the identity operator, at the initial time $t_{0}$. In this paper, we take the reduced Planck constant $\hbar=1$, denote $\Omega ^{\dagger}$ as the adjoint of an operator $\Omega$, and assume that all the kets are in a separable Hilbert space. \cite{Royden,Kreyszig} Let $\Vert | \alpha \rangle \Vert \equiv \langle \alpha | \alpha \rangle ^{1/2}$ for any $| \alpha \rangle$, and we require that $lim _{ t _{2} \rightarrow t_{1} } \Vert [ U( t_{2} ) -U( t _{1} ) ] | \beta \rangle \Vert = lim _{ t _{2} \rightarrow t_{1} } \Vert [ U ^{\dagger} (t_{2}) - U ^{\dagger} ( t _{1} ) ] | \beta \rangle \Vert=0$ for all $|\beta \rangle$. We take the Hamiltonian $H(t)= H_{0} + V(t)$ with two self-adjoint operators $H_{0}$ and $V(t)$ as the (time-independent) unperturbed and (time-dependent) perturbed parts. In addition, assume that $H_{0}$ can be diagonalized by orthornormal complete basis such that each eigenket satisfying
\begin{eqnarray}
H _{0} | n \rangle = E _{n} | n \rangle  
\end{eqnarray}
and $\Vert |n \rangle \Vert =1 $ can be parametrized by a positive integer $n$. Here each $E _{n}$ is an eigenvalue. For quantum optics,  \cite{Ohtsuki,Louisell} the relaxation term is composed of the pure-dephasing term 
\begin{eqnarray}
R _{Mp}( \rho (t)) = - \sum _{ ( n ^{\prime} n) } \gamma _{ n n ^{\prime} } | n \rangle \langle n |\rho (t) | n ^{\prime} \rangle \langle n ^{\prime} | 
\end{eqnarray}
and the transition term 
\begin{eqnarray}
R _{Mt} ( \rho (t) ) = - \frac{1}{2} \sum_{ ( n ^{\prime} n ) } w _{ n n^{\prime} } \{ \rho (t) , | n ^{\prime} \rangle \langle n ^{\prime} | \} + \sum _{ ( n ^{\prime} n ) } w _{ n n ^{\prime} } | n\rangle \langle n ^{\prime} | \rho (t) | n ^{\prime} \rangle \langle n|. 
\end{eqnarray}
Here each orbital in Eqs. (4) and (5) is a (normalized) eigenket of $H _{0}$ introduced by Eq. (3), $w _{ n n ^{\prime} }$ is the nonnegative coefficient for the transition from ket $| n ^{ \prime } \rangle $ to another ket $| n \rangle$ if $n \not= n ^{\prime} $, $\gamma _{n n ^{\prime} }$ is the nonnegative coefficient for the pure-dephasing rate of $\langle n | \rho (t)| n ^{\prime} \rangle $ when $n \not= n ^{\prime}$, and we denote $\{ A, B \} \equiv AB+BA$ for any two operators $A$ and $B$. For convenience, we take $w _{ n n } = \gamma _{n n } =0$ for all integers $n$ in this paper, so we do not need to set the condition $n \not= n ^{\prime} $ in the summations over $( n^{\prime} n )$ in Eqs. (4) and (5). If there is no pure-dephasing term, we can reduce the Markoff master equation as 
\begin{eqnarray}
\frac{\partial}{\partial t} \rho (t) = i[ \rho (t), H(t)] + R _{Mt} ( \rho (t)) \text{ \ \ \ \ \ \ \ \ \ \ \ \ \ \ \ \ \ \ \ \ \ \ \ \ \ \ \ \ \ \ \ \ \ \ \ \ \ \ \ \ \ \ \ \ \ \ \ \ } 
\end{eqnarray}
\[
\text{ \ \ \ \ \ } = i[ \rho (t), H(t)]- \frac{1}{2} \sum _{ ( n ^{\prime} n ) } w _{ n n ^{\prime} } \{ \rho (t) , | n ^{\prime} \rangle \langle n ^{\prime} | \} + \sum _{ ( n ^{\prime} n ) } w _{ n n^{\prime} } | n \rangle \langle n ^{\prime} | \rho (t) | n ^{\prime} \rangle \langle n|.
\]
On the other hand, the famous relaxation term of Lindblad form \cite{Kampen,Lindblad}
\begin{eqnarray}
R_{L}(\rho (t))=-\frac{1}{2} \sum_{l} \{ \rho (t), {\cal V} _{l} {\cal V} _{l} ^{ {\dagger} } \} + \sum_{l} {\cal V}_{l} 
^{ {\dagger } } \rho (t) {\cal V} _{l} 
\end{eqnarray}
has been derived mathematically, where $\{ {\cal V} _{l} \}$ is a set of operators. By setting 
\begin{eqnarray}
l=( n ^{\prime} n ) \text{ and } {\cal V} _{l} = w _{ n n^{\prime} } ^{1/2} | n ^{\prime} \rangle \langle n|,  
\end{eqnarray}
we can re-obtain $R _{Mt} ( \rho (t) )$ from $R _{L} ( \rho (t) )$. Therefore, Eq. (6) is a particular Markoff master equation of Lindblad form. 

The relaxation term $R_{Mt}(\rho (t))$ in Eq. (6), in fact, is composed of the loss factor 
\begin{eqnarray}
L _{M} ( \rho (t)) = - \frac{1}{2} \sum _{ ( n ^{\prime} n ) } w _{ n n^{\prime} } \{ \rho (t), | n ^{\prime} \rangle \langle n^{\prime } | \} 
\end{eqnarray}
and the gain factor 
\begin{eqnarray}
G _{M} ( \rho (t)) = \sum_{ ( n ^{\prime} n ) } w _{ n n ^{\prime} } | n \rangle \langle n ^{\prime} | \rho (t) | n 
^{\prime} \rangle \langle n |. 
\end{eqnarray}
With some calculations, we can see that the loss (gain) factor is to decrease (increase) the number of particles. Define the  function $f(n,t) \equiv \langle n|\rho (t)|n\rangle$, the loss and gain rates are $\sum _{ n^{\prime} } w _{ n ^{\prime} n } f(n ,t)$ and $\sum _{ n^{\prime} } w _{ n n ^{\prime} } f(n ^{\prime},t)$ for each orbital $|n\rangle$, respectively. \cite{Louisell} In addition, from Eq. (6) we have 
\begin{eqnarray}
\frac{\partial}{\partial t} \langle n | \rho (t) | n ^{\prime} \rangle = i(E_{n^{\prime }}-E_{n})\langle n | \rho (t) | n ^{\prime} \rangle - \frac{1}{2} \sum _{ m } ( w _{ m n ^{\prime} } + w _{ m n } ) \langle n | \rho (t)| n ^{\prime} \rangle 
\end{eqnarray}
for all phases $\langle n | \rho (t) | n ^{\prime} \rangle $ with $ n \not= n ^{\prime}$ if the perturbed potential $V(t)=0$. Thus the Markoff master equation can incorporate decoherent effects. \cite{Blum,Louisell} The decoherence comes from the loss factor $L _{M} (\rho (t))$. From Cauchy-Schwartz inequality, we have 
\begin{eqnarray}
| \langle n | \rho (t) | n ^{\prime} \rangle | \leq  \langle n | \rho (t) | n \rangle ^{1/2} \langle n ^{\prime} | \rho (t) | n ^{\prime} \rangle ^{1/2}.
\end{eqnarray}
If $\langle n | \rho (t) | n \rangle  \rightarrow 0 $ because of the loss effects on orbital $| n \rangle$, the phase $ \langle n | \rho (t) | n ^{\prime} \rangle $ should also approach zero for any other $ n ^{\prime}$ based on the above equation. Since Cauchy-Schwartz inequality is valid for all positive matrices, it is natural that the loss term results in the decoherence to preserve the positivity of $\rho (t)$. When $H(t)=H_{0}$, actually the Pauli master equation \cite{Blum,Louisell,Liboff} 
\begin{eqnarray}
\frac{\partial}{\partial t} f(n,t) = - \sum _{ n ^{\prime} } w _{ n^{\prime} n } f(n,t) + \sum _{ n^{\prime } } w _{ n n^{\prime} } f( n ^{\prime} , t) 
\end{eqnarray}
can be obtained from Eq. (6). 

For a system composed of noninteracting identical fermions, usually the matrix $ \rho (t)$ is taken as the one-particle density matrix with the trace $tr \rho (t) = \sum _{n} \langle n | \rho (t) | n \rangle = \sum _{n} f(n,t)$ equals the number of particles. \cite{Huang,Geoke} In such a case, the function $f(n,t)=\langle n | \rho (t) |n \rangle$ is interpreted as the occupation number in orbital $|n\rangle $ at time $t$. It is known that Eq. (13) should be modified for fermions to follow Pauli's exclusion principle. \cite{Liboff} By extending the coefficient $w _{ n n^{\prime} } $ as a time-dependent function 
\begin{eqnarray}
w_{nn^{\prime }}(t)=\omega _{ n n ^{ \prime } } [ 1-f(n,t) ], 
\end{eqnarray}
the master equation of the following form \cite{Iotti,Huang,Ralph,Liboff,Singh} 
\begin{eqnarray}
\frac{\partial}{\partial t}f(n,t) = -\sum _{ n ^{\prime} } \omega _{ n^{\prime} n} [ 1 - f(n ^{\prime} , t ) ] f(n,t) + \sum _{ n ^{\prime} } \omega _{ n n^{\prime} } [ 1 - f(n,t) ] f(n ^{\prime} , t ) 
\end{eqnarray}
can be obtained for fermions from Eq. (13). Here $\omega _{ n n^{\prime} }$ is the nonnegative coefficient for the transition from $n^{\prime}$ to $n$ when $n \not= n ^{\prime}$, and we set $\omega _{ n n }=0$ for all $n$. On the other hand, the nonlinear quantum master equation 
\begin{eqnarray} 
\frac{\partial }{\partial t}\rho (t) = i[\rho (t),H(t)]-\frac{1}{2} \sum _{ ( n ^{\prime} n ) }\omega _{ n n^{\prime} } ( 1 - \langle n | \rho (t) | n \rangle ) \{ \rho (t), | n ^{\prime}\rangle \langle n^{\prime } | \} 
\end{eqnarray}
\[
+ \frac{1}{2} \sum _{ ( n ^{\prime} n ) } \omega _{ n n^{\prime} } \langle n ^{\prime} | \rho (t) | n ^{ \prime } \rangle \{ I - \rho (t) , | n \rangle \langle n | \}
\]
is introduced recently for systems composed of noninteracting identical fermions. \cite{Huang} The operator 
\begin{eqnarray}
\rho ^{ ( \overline{p} ) } (t) \equiv I - \rho (t),
\end{eqnarray}
which appears in the last term of Eq. (16), can represent holes actually. \cite{Huang,Cui} In Ref. \cite{Huang}, Eq. (16) is obtained from the following equation   
\begin{eqnarray}
\frac{\partial}{\partial t} \rho (t) = i[ \rho (t), H(t)] + \{ \rho (t) , A_{p}(t) \} - \{ I- \rho (t) , A _{ \overline{p} } (t) \} 
\end{eqnarray}
by considering the conservation of the number of particles in each transition. If we define 
\begin{eqnarray}
A_{1}( \Omega ) \equiv \frac{1}{2} \sum _{ ( n ^{\prime} n ) } \omega _{n n ^{\prime} }( 1 - \langle n | \Omega | n \rangle ) | n ^{\prime} \rangle \langle n ^{\prime} | 
\end{eqnarray}
\[
A _{2}( \Omega ) \equiv \frac{1}{2} \sum _{ ( n ^{\prime} n) } \omega _{ n n ^{\prime} } \langle n ^{\prime}| \Omega | n ^{\prime} \rangle |n \rangle \langle n|, 
\]
for any operator $\Omega$, Eq. (16) can be obtained from Eq. (18) by setting $A _{p} (t)  = - A _{1} (\rho (t))$ and $A _{ \overline{p} } (t) = -A _{2} ( \rho (t))$. We can see that Eq. (16) is equivalent to the one-body master equation introduced by Ref. \cite{Ralph} after expanding $\rho (t)$ with respect to the eigenorbitals of $H_{0}$. Such an equation is denoted as Kohn-Sham master equation in Ref. \cite{Burke} because it can be used to extend Kohn-Sham equations. When $H(t)=H_{0}$, Eq. (15) can be derived from Eq. (16). \cite{Huang,Ralph} On the other hand, we can reduce Eq. (16) to Eq. (6) in the low-density limit by setting $ \omega _{ n n ^{\prime} } = w _{ n n ^{\prime} } $. \cite{Huang} Therefore, Eq. (16) can be used to unify Eqs. (6) and (15).
     
The second term at the right hand side of Eq. (16) 
\begin{eqnarray} 
L _{f} ( \rho (t) ) \equiv - \frac{1}{2} \sum _{ ( n^{\prime} n ) } \omega  _{n n ^{\prime} } ( 1 - \langle n | \rho (t) | n \rangle ) \{ \rho (t) , | n ^{\prime} \rangle \langle n ^{\prime} | \} = - \{ \rho (t), A _{1} (\rho (t)) \}, 
\end{eqnarray}
in fact, can be obtained from the loss factor $L _{M}( \rho (t))$ of Eq. (6) by Eq. (14). With some calculations, it is easy to see that $L _{f} ( \rho (t) ) $ induces the loss of particles just as $ L _{M} ( \rho (t) ) $ does. Hence $L_{f} (\rho (t))$ serves as the loss factor in Eq. (16). For each orbital $|n\rangle$, the loss rate due to $L _{f} (\rho (t))$ equals $\sum _{ n ^{\prime} } \omega _{ n ^{\prime} n } [ 1-f( n ^{\prime},t) ] f(n,t) $. In addition, the factor $L _{f} (t)$ also results in the decoherence, which is important to preserve the positivity of $\rho (t)$ as mentioned above. On the other hand, the third term of Eq. (16)
\begin{eqnarray}
G _{f} (\rho (t)) = \frac{1}{2} \sum _{ ( n ^{\prime} n) } \omega _{n n^{\prime}} \langle n ^{\prime} | \rho (t) | n ^{\prime} \rangle \{ I - \rho (t), | n \rangle \langle n | \} = \{ I - \rho (t), A _{2} ( \rho (t) )\}
\end{eqnarray}
induces the gain of particles and serves as the gain factor. For each orbital $|n\rangle$, the gain rate due to $G _{f} (\rho (t) )$ equals $\sum _{ n^{\prime} } \omega _{ n n^{\prime} } [ 1-f(n,t) ] f(n ^{\prime},t)$. Such a factor, however, cannot be obtained from the gain factor $G _{M} ( \rho (t)) $ of Eq. (6) by Eq. (14). With some calculations, we note that $G _{f} (\rho (t))$ results in the decoherence while $G _{M} (\rho (t))$ does not. 

The gain of particles (holes), in fact, is just the loss of holes (particles) in each specific orbital. \cite{Huang} Hence it is natural to relate the the gain factor $G _{f} ( \rho (t))$, which can be taken as the loss factor for holes, to $L _{f} (\rho (t) )$. We note   
\begin{eqnarray}
L _{f} ( \rho (t)) = -\frac{1}{2} \sum _{ ( n ^{\prime} n )} \omega _{ n n ^{\prime} } \langle n | \rho ^{ 
( \overline{p} ) } (t) | n \rangle \{ I - \rho ^{ ( \overline{p} ) } (t) , | n ^{\prime} \rangle \langle n ^{\prime} | \}, 
\end{eqnarray}
from which we can obtain the gain factor $G _{f} (\rho (t))$ by  $ \omega _{n n ^{\prime} } \rightarrow \omega _{ n ^{\prime} n }$ , changing the sign, and relacing $\rho ^{ ( \overline{p} ) } (t)$ by $\rho (t)$. The process that particles jump from $n$ to $n^{\prime }$ should correspond to the transition of holes from $n^{\prime}$ to $n$, and thus it is reasonable to replace $\omega _{n n ^{\prime} }$ by $ \omega _{ n ^{\prime} n }$ to construct $G _{f} ( \rho (t))$ from $L _{f}(\rho (t))$. The change of the sign is natural because the loss factor is to decrease the occupation number while the gain factor is to increase it. We shall replace $ \rho ^{ (\overline{p}) } ( t ) $ by $ \rho (t)$ since $ L _{f} ( \rho (t) )$ and $G _{f} ( \rho (t))$ correspond to the gain of holes and that of particles, respectively. 

As mentioned above, the decoherence due to the loss factor $L _{f} ( \rho (t) ) $ (for particles) is important to preserve the positivity of $\rho (t)$. Since $G _{f} ( \rho (t))$ can be taken as the loss factor for holes, it is natural for $G _{f} ( \rho (t))$ to result in the decoherence to preserve the positivity of $\rho ^{ ( \overline{p} ) } (t)$. The positivity of $\rho ^{ ( \overline{p} ) } (t)$, in fact, is equivalent to that $\rho (t)$ follows Pauli's exclusion principle because $ \langle \alpha | \rho (t) | \alpha \rangle \leq  1 $ iff
\begin{eqnarray}
 \langle \alpha | \rho ^{ ( \overline{p} ) } (t) | \alpha \rangle = 1 - \langle \alpha | \rho (t) | \alpha \rangle \geq  0 
\end{eqnarray} 
for any normalized $| \alpha \rangle$. Therefore, the decoherence is important not only to the positivity, but also to Pauli's exclusion principle.         

\section{The positivity and Pauli's exclusion principle}

To show that Eq. (16) is suitable for fermions, it will be proven in this section that both the positivity and Pauli's exclusion principle are preserved when there exists a positive real number $M$ such that both  
\begin{eqnarray}
\sup _{n} \sum _{ n ^{\prime} } \frac{1}{2} w _{ n n ^{\prime} } \text{ and } \sup _{n} \sum _{ n^{\prime} } \frac{1}{2} w _{ n^{\prime} n }  \leq M.
\end{eqnarray}
For a specific $ | n \rangle $, the loss and gain rates $\sum _{ n ^{\prime} } w _{ n^{\prime} n } [ 1 - f( n ^{ \prime } (t) ] f(n,t)$ and $ \sum _{ n ^{\prime} } w _{ n n ^{\prime} } [ 1 - f(n,t) ] f ( n ^{\prime} ,t )$ should be both smaller than $2M$. So such a number provides an upper bound for the transition rate. For convenience, let $t _{0}$ be the initial time for the time evolution of $\rho (t)$. Just as mentioned in the last section, the unitary operator $U(t)$ generated by the Hamiltonian $H(t)$ equals the identity operator $I$ at $t=t _{0}$. We require $\rho (t_{0})$ is a positive self-adjoint operator following Pauli's exclusion principle such that 
\begin{eqnarray}
0 \leq \langle \alpha | \rho ( t _{0} ) | \alpha \rangle \leq 1 \text{ for all normalized } | \alpha \rangle.
\end{eqnarray}
Because Eq. (16) yields an initial-value problem, we just need to prove the existence and uiqueness of the solution when $t$ is in a time interval $[t _{0}, t_{f}]$ for some final time $t_{f}$ satisfying
\begin{eqnarray}
t_{0} < t _{f} < t _{0} + \frac{1}{4M}. 
\end{eqnarray}
In the following, the notation $||Q|| \equiv \sup_{ \Vert  | \alpha \rangle \Vert = 1 } \Vert  Q | \alpha \rangle \Vert $ as the natural norm of any bounded (linear) operator $Q$ \cite{Royden,Kreyszig}. It is known that $||Q ^{\dagger}||= ||Q||$ under such a norm. \cite{Kreyszig} In addition, we denote $|||{\cal Q}(t)||| \equiv \sup _{ t \in [t _{0},t _{f} ] } ||{\cal Q} (t)||$ for any time-dependent operator ${\cal Q} (t)$ if $ \sup _{ t \in [t_{0},t_{f}] } ||{\cal Q} (t)||$ is finite.    

It is convenient to introduce the following Banach spaces \cite{Royden,Kreyszig,textbook} ${\cal S}_{1}$ and ${\cal S} ^{ \prime } _{1}$, a subset ${\cal S} _{2}$ of ${\cal S} _{1}$, and a vector space ${\cal S} _{0}$: \newline
{\bf Definition 3.1} $Let$ ${\cal S} ^{\prime} _{1}$ $be$ $the$ $Banach$ $space$ $composed$ $of$ $all$ $the$ $mappings$ $\Omega ^{\prime} (t)$ $from$ $the$ $time$ $interval$ $[t_{0},t_{f}]$ $to$ $bounded$ ($linear$) $operators$ $such$ $that$ 
\begin{eqnarray}
\text{ \ } \lim _{ t _{2} \rightarrow t _{1} } || \Omega ^{\prime} (t_{2} ) - \Omega ^{\prime} (t_{1}) || =0 , 
\end{eqnarray}
$and$ $denote$ the $Banach$ $space$ ${\cal S} _{1}=\{ \Omega (t) | \Omega (t) = U (t) \Omega ^{\prime} (t) U ^{\dagger} (t) \text{ for some } \Omega ^{\prime} (t) \in {\cal S} ^{\prime} _{1} \}$. $We$ $take$ $|||\cdot||| \text{ } as \text{ } the \text{ } norm \text{ } on \text{ } {\cal S} _{1}$ $and$ ${\cal S} ^{ \prime } _{1}$. $The$ $set$ ${\cal S} _{2} \equiv \{ \Omega (t) \in {\cal S} _{1}| \Omega (t) = \Omega ^{ \dagger } (t), 0 \leq \langle \alpha | \Omega (t) | \alpha \rangle \leq 1 \text{ }for \text{ } all \text{ } normalized \text{ } | \alpha \rangle \}$ $is$ $a$ $complete$ $subset$ $of$ ${\cal S} _{1}$. $The$ $vector$ $space$ ${\cal S} _{0}$ $is$ $composed$ $of$ $all$ $the$ $mappings$ ${\cal Q} (t)$ from $[t _{0} ,t _{f}]$ $to$ $bounded$ $operators$ $such$ $that$   
\begin{eqnarray}
||| {\cal Q} (t)||| < \infty \text{ and } \lim _{ t_{2} \rightarrow t _{1} } \Vert [ {\cal Q} (t_{2}) - {\cal Q} ( t_{1} ) ] | \alpha \rangle \Vert = 0 \text { for all } | \alpha \rangle .
\end{eqnarray}    
\newline

The unitary operator $U(t)$ and its adjoint $U ^{\dagger} (t)$ are in ${\cal S} _{0}$. A density matrix $\rho (t)$ (in the Schr\"{o}dinger picture) is a postive one following Pauli's exclusion principle if $ \rho (t) \in {\cal S} _{2}$. For each operator $\Omega (t) \in {\cal S} _{1}$ in the Schr\"{o}dinger picture, its corresponding operator in the Heisenberg picture is $ \Omega ^{\prime} (t) \equiv U ^{\dagger}(t) \Omega (t) U(t) \in {\cal S} ^{\prime} _{1}$. The adjoint of any operator in ${\cal S} ^{\prime} _{1}$ is also in ${\cal S} ^{\prime} _{1}$. Actually ${\cal S} _{1}$ and ${\cal S} ^{\prime} _{1}$ are subspaces of ${\cal S} _{0}$. For convenience, any operator in ${\cal S} ^{\prime} _{1}$ is denoted by the superscript $\lq \lq \prime "$ in this paper. As discussed in Appendix A, we can denote the operator $ {\cal O} = \int _{ t_{1} } ^{ t_{2} } d t ^{\prime} {\cal Q} ( t ^{\prime} ) $ for any ${\cal Q} (t) \in {\cal S} _{0}$ iff ${\cal O}$ is the uique one following $\langle \alpha |{\cal O} | \beta \rangle = \int _{ t_{1} } ^{ t_{2} } d t ^{\prime} \langle \alpha | {\cal Q} ( t ^{\prime} ) | \beta \rangle$ for all $| \alpha \rangle$ and $ | \beta \rangle$. Here $t _{1}$ and $t_{2}$ follow $t _{0} \leq t_{1} \leq t_{2} \leq t_{f}$. In addition, we have 
\begin{eqnarray}\
\Vert {\cal O} \Vert = \Vert \int _{ t_{1} } ^{ t_{2} } d t ^{\prime} {\cal Q} ( t ^{\prime} ) \Vert \leq |t_{2}-t_{1}| \times |||{\cal Q} (t)|||.
\end{eqnarray} 
For any time-dependent operator ${\cal Q} (t) \in {\cal S} _{0}$, we denote
\begin{eqnarray}
\frac{\partial}{\partial t} \Omega ^{\prime} (t) = {\cal Q} (t) \text{ for some } \Omega ^{\prime} (t) \in {\cal S} ^{\prime} _{1}
\end{eqnarray}
on an interval $[t_{1},t_{2}] \subseteq [t_{0},t_{f}]$ iff 
\begin{eqnarray}
\Omega ^{\prime} (t) = \Omega ^{\prime} ( t _{1} ) + \int _{ t _{1} } ^{t} dt ^{\prime} {\cal Q} ( t ^{\prime} )
\end{eqnarray}
at any $t \in [t_{1},t_{2}]$. It should be noted that such a definition can be invalid for $\frac{\partial}{\partial t} \Omega(t)$ when $\Omega (t) \in {\cal S} _{1}$. Some properties about the integral and drivative of operators are discussed in Appendix A.

The following lemma is important to the proof:\newline
{\bf Lemma 3.2} $Consider$ $the$ $equation$ 
\begin{eqnarray}
\frac{\partial}{\partial t} \rho (t) = i [ \rho (t), H(t) ] - \{ \rho (t) , {\cal A} (t) \} + {\cal B} (t) \text{ as } t \in [t_{0},t_{f}], 
\end{eqnarray}   
$and$ $assume$ $that$ $the$ $given$ $initial$ $matrix$ $\rho (t _{0})$ $satisfies$ $Eq.$ $(25)$. $Here$ ${\cal A} (t)$ $and$ ${\cal B} (t)$ $are$ $two$ $positive$ $self$-$adjoint$ $operators$ $in$ ${\cal S} _{0}$, $and$ $we$ $assume$ $that$ $|||{\cal A} (t)||| \leq 2M$. $Then$ $there$ $exists$ $a$ $unique$ $self$-$adjoint$ $matrix$ $\rho (t) \in {\cal S} _{1}$ $to$ $follow$ $the$ $above$ $equation$, $and$ $the$ $positivity$ $is$ $preserved.$  \newline
{\bf proof} $\text{ \ }$ To avoid unbounded problems due to the Hamiltonian $H(t)$, we can rewrite Eq. (32) as   
\begin{eqnarray}
\frac{\partial}{\partial t} \rho ^{ \prime } (t) =  - \{ \rho ^{ \prime } (t) , {\cal A} _{U} (t)  \} +  {\cal B} _{U} (t)
\end{eqnarray}
by introducing $\rho ^{\prime} (t) \equiv U ^{\dagger} (t) \rho (t) U(t)$ as the density matrix in the Heisenberg picture. Here ${\cal A} _{U} (t) \equiv U ^{\dagger} (t) {\cal A} (t) U(t)$ and ${\cal B} _{U} (t) \equiv U ^{\dagger} (t) {\cal B} (t) U(t)$. We have $\rho ^{\prime} (t _{0}) = \rho (t _{0} )$ at the initial time $t _{0}$ because $U(t _{0})=I$. The matrix $\rho ^{\prime} (t) \in {\cal S} ^{\prime} _{1}$ iff $\rho (t) \in {\cal S} _{1}$, and we can prove that both ${\cal A} _{U} (t)$ and ${\cal B} _{U} (t) \in {\cal S} _{0}$ from corollary A1. Let ${\cal F}$ be the mapping from ${\cal S} _{1} ^{\prime}$ to ${\cal S} _{1} ^{\prime}$ such that $\Lambda ^{\prime} (t) = {\cal F} ( \Omega ^{\prime} (t) )$ iff
\begin{eqnarray}
\Lambda ^{\prime} (t) =  \rho ( t_{0} ) + \int _{ t _{0} } ^{t} dt ^{ \prime } (  {\cal B} _{U} (t ^{\prime}) - \{ \Omega ^{ \prime } (t ^{\prime}) , {\cal A} _{U} (t ^{\prime}) \}).
\end{eqnarray}
The mapping ${\cal F}$ is well-defined and bounded because we can use corollary A1 to prove that the integrand $  {\cal B} _{U} (t) - \{ \Omega ^{ \prime } (t) , {\cal A} _{U} (t) \} \in {\cal S} _{0}$. In addition, we can see the equivalence between Eq. (33) and $\rho ^{\prime} (t) = {\cal F} ( \rho ^{\prime} (t) )$. Since $|||{\cal A} _{U} (t)||| =|||{\cal A} (t)||| \leq 2M$, ${\cal F}$ is a contraction on ${\cal S} ^{\prime} _{1}$ and there exists a unique solution $\rho ^{\prime} (t) \in {\cal S} _{1} ^{\prime}$ to Eq. (33) from the fixed point theory. \cite{textbook} We can take Eqs. (32) and (33) as the same equation in different pictures, so $\rho (t) = U(t) \rho ^{\prime} (t) U ^{\dagger} (t)$ is the unique solution to Eq. (32). 

Define a mapping $ K ^{ \prime} (t;t ^{\prime}) $ from $t ^{\prime} \in [t_{0},t _{f}]$ to ${\cal S} ^{\prime} _{1}$ by
\begin{eqnarray}
\frac{\partial}{\partial t} K ^{ \prime} ( t ; t ^{\prime} ) = - {\cal A} _{U} (t) K ^{\prime} (t;t^{\prime}) \text{ at } t \geq t^{\prime} 
\end{eqnarray}
and $ K ^{ \prime} (t ; t^{\prime}) = I \text{ at } t \leq t ^{\prime}$. In addition, let $K (t;t^{\prime})$ be a two-parameter operator such that $ K (t;t^{\prime}) = U (t) K ^{\prime} (t;t^{\prime}) U ^{\dagger} (t ^{\prime})$ when $t > t ^{\prime}$ and $K (t;t^{\prime}) = I $ when $t \leq t ^{\prime}$. It is shown in Appendix B that $ K ^{ \prime} (t;t^{\prime})$ can also be taken as a mapping from $t \in [t_{0},t_{f}]$ to ${\cal S} ^{\prime} _{1}$, and we have $\frac{\partial}{\partial t} K ^{ \prime \dagger} (t;t^{\prime})= - K ^{ \prime \dagger} (t;t^{\prime}) {\cal A} _{U} (t)$ at $t \geq t ^{\prime}$ from corollary A5. On the other hand, $\frac{\partial}{\partial t} K (t;t^{\prime})$ may suffer unbounded problems if $H (t)$ is not bounded although it is convenient to introduce the following equation
\begin{eqnarray}
i\frac{\partial}{\partial t} K (t;t^{\prime}) = [ H (t) - i {\cal A} (t) ] K (t;t^{\prime}) \text{ at } t \geq t ^{\prime}.  
\end{eqnarray}
To prove the self-adjointness and positivity, we note that the solutions to Eqs. (33) and (32) are  
\begin{eqnarray}
\rho ^{ \prime } (t) = K ^{\prime} (t;t_{0}) \rho ( t _{0}) K ^{\prime \dagger} (t;t_{0}) + \int _{t _{0}} ^{t} dt ^{\prime} K ^{\prime} (t;t^{\prime}) {\cal B} _{U} (t ^{\prime}) K ^{ \prime \dagger} (t;t^{\prime})
\end{eqnarray}
\begin{eqnarray}
\rho (t) = K (t;t_{0}) \rho ( t _{0}) K ^{\dagger} (t;t_{0}) + \int _{t _{0}} ^{t} dt ^{\prime} K (t;t^{\prime}) {\cal B} (t ^{\prime}) K ^{\dagger} (t;t^{\prime}),
\end{eqnarray}
respectively. It is discussed in Appendix B how to prove that both Eqs. (37) and (38) are well-defined. The above two equations can be related by $\rho (t) = U (t) \rho ^{\prime} (t) U ^{\dagger} (t)$, and we just need to check $\rho ^{\prime} (t _{0} )$ and $\frac{\partial}{\partial t} \rho ^{\prime} (t)$ to see that they provide the solutions. We can use corollaries A2 and A4 to perform the time-derivative on $ \rho ^{\prime} (t)$. It is easy to see that $\rho (t)$ and $\rho ^{\prime} (t)$ constructed by the above two equations are self-adjoint. To prove that $\rho (t)$ is positive such that $\langle \alpha | \rho (t) | \alpha \rangle \geq 0$ for all $|\alpha\rangle$, we note that 
\begin{eqnarray}
\langle \alpha | \rho (t) | \alpha \rangle = \langle \alpha (t,t_{0}) | \rho ( t _{0}) | \alpha ( t , t _{0} ) \rangle + \int _{t _{0}} ^{t} dt ^{\prime} \langle \alpha (t,t ^{\prime}) | {\cal B} ( t ^{\prime} ) | \alpha (t,t ^{\prime}) \rangle \geq 0
\end{eqnarray} 
from Eq. (38) if we define $|\alpha (t _{1},t _{2}) \rangle \equiv K ^{\dagger} ( t _{1};t_{2}) | \alpha \rangle$. $\text{ \ \ \ \ \ \ \ \ \ \ \ \ \ \ \ \ \ \ \ \ \ \ \ \ \ \ \ \ \ \ \ \ \ \ \ \ \ \ \ \ \ }$ {\bf QED} \newline 

It is mentioned in the last section that Eq. (16) can be obtained from Eq. (18) by setting $A _{p} (t) = - A _{1} (\rho (t))$ and $A _{ \overline{p} } (t) = - A _{2} ( \rho (t))$. Thus it is convenient to discuss Eq. (18) before completing the proof for Eq. (16). Based on lemma 3.2, it is proven in the following proposition that Eq. (18) preserves both the positivity and Pauli's exclusion principle. \newline
{\bf Proposition 3.3} $\text{ \ }$ $Let$ $A _{p} (t)$ $and$ $A _{ \overline{p} } (t)$ $be$ $two$ $time$-$dependent$ $self$-$adjoint$ $operators$ $in$ ${\cal S} _{0}$. $Assume$ $that$ $both$ $-A _{p} (t)$ $and$ $-A _{ \overline{p} } (t)$ $are$ $positive$ $and$ $|||A _{p} (t) + A _{ \overline{p} } (t) ||| \leq 2M$. $Then$ $there$ $exists$ $a$ $unique$ $matrix$ $\rho (t) \in {\cal S} _{2}$ $satisfying$ $Eq.$ $(18)$ $if$ $Eq.$ $(25)$ $is$ $valid$. \newline
{\bf proof} $\text{ \ }$ We can rewrite Eq. (18) as
\begin{eqnarray}
\frac{\partial}{\partial t} \rho (t) = i [ \rho (t) , H(t) ] + \{ \rho (t) , A _{p} (t) + A _{ \overline{p} } (t) \} - 2 A _{ \overline{p} } (t). 
\end{eqnarray}
By setting ${\cal A}(t)= - A _{p} (t) - A _{ \overline{p} } (t)$ and ${\cal B}(t)= - 2 A _{ \overline{p} } (t)$, the above equation is of the same form as that of Eq. (32) and the condition $|||{\cal A} (t)||| \leq 2M $ is satisfied. Hence there exists a unique self-adjoint solution $\rho (t) \in {\cal S} _{1}$ following 
\begin{eqnarray}
\langle \alpha | \rho (t) | \alpha \rangle \geq 0
\end{eqnarray}
for all normalized $| \alpha \rangle$ from lemma 3.2. On the other hand, we can also rewrite Eq. (18) with respect to $\rho ^{ ( \overline{p} ) } (t) = I - \rho (t) $ as
\begin{eqnarray}
\frac{\partial}{\partial t} \rho ^{ ( \overline{p} ) } (t) = i [ \rho ^{ ( \overline{p} ) } (t) , H(t) ] + \{ \rho ^{ ( \overline{p} ) } (t) , A _{p} (t) + A _{ \overline{p} } (t) \} - 2 A _{ p } (t)
\end{eqnarray}       
The above equation is also of the form of Eq. (32), and Eq. (25) indicates that $\rho ^{ ( \overline{p} ) } (t) $ is positive. Then we have 
\begin{eqnarray} 
1 - \langle \alpha | \rho (t) | \alpha \rangle = \langle \alpha | \rho ^{ ( \overline{p} ) } (t) | \alpha \rangle \geq 0 
\end{eqnarray}
for all normalized $| \alpha \rangle$. We can complete the proof from Eqs. (41) and (43). $\text{ \ \ \ \ \ \ \ \ \ }$ {\bf QED} \newline  

Just as in lemma 3.2, we can rewrite Eq. (40) with respect to $ \rho ^{\prime} (t) \equiv U ^{\dagger} (t) \rho (t) U (t)$ to avoid the unobunded problem due to $H(t)$. In the Heisenburg picture, the density matrices for particles and holes are $\rho ^{\prime} (t)$ and $ I -  \rho ^{\prime} (t) = U ^{\dagger} (t) \rho ^{( \overline{p} )} (t) U (t)$, respectively. We can see the meaning of $\frac{\partial}{\partial t} \rho ^{( \overline{p} )} $, which appears in Eq. (42), by considering the time-derivative on $I -  \rho ^{\prime} (t) $.       

Now we return to discuss Eq. (16). As shown in Appendix C, both $A _{1} (\Omega (t))$ and $A _{2} ( \Omega (t) )$ are positive self-adjoint operators in ${\cal S} _{0}$ for any $\Omega (t) \in {\cal S} _{2}$. In addition, direct calculations yield that  
\begin{eqnarray}
|||A _{j}( \Omega _{1} (t) ) - A _{j} ( \Omega _{2} (t) ) ||| \leq M ||| \Omega _{1} (t) - \Omega _{2} (t) |||
\end{eqnarray}
for any two operators $\Omega _{1} (t)$ and $\Omega _{2} (t)$ in ${\cal S} _{2}$, where $j=1$ and $2$. The solution to Eq. (16), in fact, satisfies $\rho (t) = F (\rho (t))$ if we define the mapping $F$ from ${\cal S} _{2}$ to ${\cal S} _{2}$ as: \newline
{\bf Definition 3.4} $\text{ \ }$ $For$ $any$ $\Omega (t) \in {\cal S} _{2},$ $let$  
\begin{eqnarray}
F (\Omega (t)) \equiv {\cal K} _{ \Omega } (t ; t_{0}) \rho ( t _{0}) {\cal K} _{\Omega} ^{\dagger} (t ; t_{0}) + 2 \int _{ t _{0} } ^{t} dt ^{\prime} {\cal K} _{\Omega} (t; t ^{\prime}) A _{2} ( \Omega ( t^{ \prime } )) {\cal K} ^{ \dagger } _{\Omega} ( t; t ^{\prime} ). 
\end{eqnarray}
$Here$ ${\cal K} _{ \Omega } (t; t ^{\prime} )$ $follows$ 
\begin{eqnarray}
i\frac{\partial}{\partial t} {\cal K} _{ \Omega } (t; t ^{\prime}) = [ H (t) - i A _{1}( \Omega (t) ) - i A _{2} ( \Omega (t) ) ] {\cal K} _{ \Omega } ( t ; t ^{\prime} ) \text{ at } t \geq t ^{\prime} 
\end{eqnarray}
$and$ ${\cal K} _{ \Omega } (t; t ^{\prime} ) = I$ $at$ $t \leq t ^{\prime}$. $Then$  
\begin{eqnarray}
\frac{\partial}{\partial t} \Lambda (t) = i [ \Lambda (t) , H (t) ] - \{ \Lambda (t) , A _{1} ( \Omega (t) ) \} + \{ I - \Lambda (t) , A _{2} (\Omega (t)) \} 
\end{eqnarray} 
\[
= i [ \Lambda (t) , H (t) ] - \{ \Lambda (t) , A _{1} ( \Omega (t) ) + A _{2} ( \Omega (t) )\} + 2 A _{2} (\Omega (t))  \text{ at } t \geq t _{0}
\]
$and$ $\Lambda (t _{0}) = \rho (t _{0})$ $iff$ $\Lambda (t) = F(\Omega (t)) $. \newline 

We can see the equivalence between Eqs. (45) and (47) by comparing them to Eqs. (38) and (32). While Eq. (46) may suffer the unbounded problem due to the Hamiltonian $H(t)$, we note that ${\cal K} _{\Omega} (t;t^{\prime})$ is just the solution to Eq. (36), which is equivalent to Eq. (35), if we take ${\cal A} (t) = A _{1}( \Omega (t)) + A _{2} ( \Omega (t) )$ and ${\cal B}(t) = 2 A _{2} (\Omega (t))$. It is easy to check that the condition $|||{\cal A} (t)||| = |||A _{1}( \Omega (t)) + A _{2} ( \Omega (t) )||| \leq 2M$ is valid for any $\Omega (t) \in {\cal S} _{2}$. By introducing ${\cal K} ^{\prime} _{\Omega} (t;t^{\prime})$ such that ${\cal K} ^{\prime} _{ \Omega } (t; t ^{\prime} ) = U ^{\dagger} (t) {\cal K} _{ \Omega } (t; t ^{\prime} ) U (t ^{\prime})$ at $t \geq t ^{\prime}$ and ${\cal K} ^{\prime} _{ \Omega } (t; t ^{\prime} ) = I$ at $t \leq t ^{\prime}$, we can rewrite Eq. (46) as 
\begin{eqnarray}
\frac{\partial}{\partial t} {\cal K} ^{\prime} _{ \Omega } (t; t ^{\prime}) = - U ^{\dagger} (t) [ A _{1}( \Omega (t) ) + A _{2} ( \Omega (t) ) ] U(t) {\cal K} ^{\prime} _{ \Omega } (t ; t ^{\prime} )   
\end{eqnarray}
when $t \geq  t ^{\prime}$. The above equation is of the form of Eq. (35), and does not suffer the unbounded problem when the Hamiltonian is not bounded. For any two operators $\Omega _{1} (t)$ and $\Omega _{2} (t) \in {\cal S} _{2}$, we have 
\begin{eqnarray}
\sup _{t,t ^{\prime}} \Vert {\cal K} _{ \Omega _{1} } (t; t ^{\prime}) - {\cal K} _{ \Omega _{2} } (t; t ^{\prime}) \Vert  = \sup _{t,t ^{\prime}} \Vert {\cal K} ^{\prime} _{ \Omega _{1} } (t; t ^{\prime}) - {\cal K} ^{\prime} _{ \Omega _{2} } (t; t ^{\prime}) \Vert \leq 2 |||\Omega _{1} (t) - \Omega _{2} (t)|||
\end{eqnarray}
because $\sup _{t \geq t ^{\prime}} \Vert {\cal K} ^{\prime} _{ \Omega _{1} } (t; t ^{\prime}) - {\cal K} ^{\prime} _{ \Omega _{2} } (t; t ^{\prime}) \Vert \leq \sup _{t \geq t ^{\prime}} \Vert \int _{ t ^{\prime} } ^{t} d t ^{\prime \prime}  U ^{\dagger} ( t ^{ \prime \prime } ) [ A _{1} ( \Omega _{1}( t ^{\prime \prime} ) ) + A _{2} ( \Omega _{1}( t ^{\prime \prime} ) ) ] U (t ^{ \prime \prime } ) [ {\cal K} _{ \Omega _{1} } ^{\prime} ( t;t ^{\prime \prime}) - {\cal K} _{ \Omega _{2} } ^{\prime} ( t;t ^{\prime \prime}) ] \Vert + \sup _{t \geq t ^{\prime}} \Vert \int _{ t ^{\prime} } ^{t} dt ^{\prime \prime} U ^{\dagger} ( t ^{ \prime \prime } ) [ A _{1} ( \Omega _{1}( t ^{\prime \prime} ) ) + A _{2} ( \Omega _{1}( t ^{\prime \prime} ) ) - A _{1} ( \Omega _{2}( t ^{\prime \prime} ) ) - A _{2} ( \Omega _{2} ( t ^{\prime \prime} ) )] U ( t ^{ \prime \prime } ) {\cal K} _{ \Omega _{2} } ^{\prime} ( t;t ^{\prime \prime}) \Vert $. Based on proposition 3.3, we can prove that $F(\Omega (t)) \in {\cal S}_{2}$ for any $\Omega (t) \in {\cal S} _{2}$. 

By the following lemma, actually $F$ is a contraction \cite{textbook} on ${\cal S} _{2}$ when the product $M|t_{f}-t_{0}|$ is small enough. \newline
{\bf Lemma 3.5} $\text{ \ }$ $We$ $can$ $find$ $a$ $positive$ $number$ $c$ $such$ $that$ 
\begin{eqnarray}
|||F(\Omega _{1} (t)) - F (\Omega _{2} (t)) ||| < c M|t_{f}-t_{0}| \times ||| \Omega _{1} (t) - \Omega _{2} (t)|||
\end{eqnarray}
$for$ $any$ $two$ $time$-$dependent$ $operators$ $\Omega _{1} (t)$ and $\Omega _{2} (t) \in {\cal S} _{2}$. 
\newline
{\bf proof} $\text{ \ }$ Let $\Lambda _{1} (t)$ and $\Lambda _{2} (t)$ be two time-dependent operators satisfying 
\begin{eqnarray}
\frac{\partial}{\partial t} \Lambda _{j} (t) = i [ \Lambda _{j} (t) , H (t) ] - \{ \Lambda _{j} (t) , A _{1} ( \Omega _{j}(t) ) \} + \{ I - \Lambda _{j}(t) , A _{2} (\Omega _{j}(t)) \} \text{ at } t \geq t _{0}
\end{eqnarray} 
and $\Lambda _{j} (t _{0}) = \rho (t _{0})$, where $j=1$ and $2$. Then $\Lambda _{j} (t) = F( \Omega _{j} (t))$, so   
\begin{eqnarray}
\Lambda _{j} (t) =  {\cal K} _{\Omega _{j}} (t;t_{0}) \rho ( t _{0}) {\cal K} ^{\dagger} _{ \Omega _{j} } (t,t_{0}) + 2 \int _{t _{0}} ^{t} dt ^{\prime} {\cal K} _{ \Omega _{j} } (t; t ^{\prime} ) A _{2} ( \Omega _{j} ( t ^{\prime} ) ) {\cal K} ^{\dagger} _{ \Omega _{j} } (t;t ^{\prime})
\end{eqnarray} 
from definition 3.4. Here ${\cal K} _{ \Omega _{j} } ( t ; t ^{\prime} )$ is introduced by Eq. (46). From the above equation, we can complete the proof by using Eqs. (29), (44) and (49) because $\Vert Y_{1} Y_{2} Y_{3} - Z_{1} Z_{2} Z_{3} \Vert \leq \Vert Y_{1} \Vert \times \Vert Y _{2} \Vert \times \Vert  Y_{3}-Z_{3} \Vert + \Vert Y_{1} \Vert \times \Vert Y_{2} - Z_{2} \Vert \times \Vert Z_{3} \Vert + \Vert Y_{1} -Z_{1} \Vert \times \Vert Z_{2} \Vert \times \Vert Z_{3} \Vert$ for any six operators $Y_{1}$, $Y_{2}$, $Y_{3}$, $Z_{1}$, $Z _{2}$, and $Z _{3}$. \text{ \ \ \ \ \ \ \ \ \ \ \ \ \ \ \ \ \ \ \ \ \ \ \ \ \ \ \ \ \ \ \ \ \ \ \ \ \ \ \ \ \ \ \ \ \ } {\bf QED} \newline

From the above discussions, the function $F$ is a contraction on ${\cal S} _{2}$ if we choose $t_{f}$ such that $M|t_{f}-t_{0}|$ is small enough. Based on the fixed point theory \cite{textbook}, there exists a unique $\rho(t) \in {\cal S} _{2}$ satisfying $\rho (t) = F(\rho (t))$ and we can complete the proof.            
         
\section{Discussion}

In the last section, a contraction $F$ is defined on ${\cal S} _{2}$ such that the unique solution to Eq. (16) satisfies $\rho (t) = F(\rho (t))$. To obtain such a solution more explicitly, we can construct a sequence $\{ \rho _{j} (t) \}$ by 
\begin{eqnarray}
\rho _{1} (t) = \rho ( t _{0} ) 
\end{eqnarray}
as $t \in [t_{0},t_{f}]$ and    
\begin{eqnarray}
\frac{\partial}{\partial t} \rho _{j+1} (t) 
=i[ \rho _{j+1} (t) , H (t) ] - \{ \rho _{j+1} (t) , A _{1} ( \rho _{j} (t) ) ) \} + \{ I -  \rho _{j+1} (t), A _{2} ( \rho _{j} (t) ) \}
\end{eqnarray}
\[
=i[ \rho _{j+1} (t) , H (t) ] - \{ \rho _{j+1} (t) , A _{1} ( \rho _{j} (t) ) + A _{2} ( \rho _{j} (t) ) \} + 2 A _{2} ( \rho _{j} (t) )
\]
with $\rho _{j+1} (t _{0})= \rho (t _{0})$ for each positive integer $ j $. Then $\rho _{j+1}(t) = F(\rho _{j} (t))$, and the sequence $\{ \rho _{j} (t) \} \subseteq {\cal S} _{2}$ converges to the solution we want.  

In definition 3.1, different continuous classes are introduced for the proof. When the unitary operator is generated by a bounded self-adjoint operator $H(t) \in {\cal S} ^{\prime} _{1}$, ${\cal S} _{1}$ and ${\cal S} ^{\prime} _{1}$ are of the same class and it is not necessary to clarify them. However, usually $H(t)$ is unbounded, and the time-derivative $\frac{\partial}{\partial t} \rho (t)$ should be defined carefully in all the quantum master equations mentioned in section II. The meaning of the time-derivative on $\rho (t)$, in fact, becomes more clear if we trasform $\rho (t)$ to the corresponding matrix $\rho ^{\prime} (t) = U ^{\dagger} (t) \rho (t) U (t)$ in the Heisenberg picture. In lemma 3.2, proposition 3.3, and definition 3.4, the density matrix $\rho (t) \in {\cal S} _{1}$ and thus $\rho ^{\prime} (t) \in {\cal S} ^{\prime} _{1}$, on which the time-derivative can be defined by Eqs. (30) and (31).        

It is shown in Ref. \cite{Huang} that Eq. (16) can be reduced to Eq. (6) when $\rho (t)$ is of the low-density distribution such that $I - \rho (t) \sim I$. By rewriting Eq. (16) with respect to $\rho ^{ ( \overline{p} ) } (t)$, we have 
\begin{eqnarray}
\frac{\partial }{\partial t}\rho ^{ ( \overline{p} ) } (t) = i[\rho ^{ ( \overline{p} ) } (t),H(t)]-\frac{1}{2} \sum _{ ( n ^{\prime} n ) }\omega _{ n^{\prime} n } ( 1 - \langle n | \rho ^{ ( \overline{p} ) } (t) | n \rangle ) \{ \rho ^{ ( \overline{p} ) } (t), | n ^{\prime} \rangle \langle n ^{\prime } | \} 
\end{eqnarray}
\[
+ \frac{1}{2} \sum _{ ( n ^{\prime} n ) } \omega _{ n^{\prime} n } \langle n ^{\prime} | \rho ^{ ( \overline{p} ) } (t) | n ^{ \prime } \rangle \{ I - \rho ^{ ( \overline{p} ) } (t) , | n \rangle \langle n | \}.
\]
When $\rho ^{ ( \overline{p} ) } (t)$ is of the low-density distribution such that $I - \rho ^{ ( \overline{p} ) } (t) \sim I$, we can reduce the above equation as 
\begin{eqnarray}
\frac{\partial }{\partial t}\rho ^{ ( \overline{p} ) } (t) = i[\rho ^{ ( \overline{p} ) } (t),H(t)]-\frac{1}{2} \sum _{  ( n ^{\prime} n ) }\omega _{ n^{\prime} n } \{ \rho ^{ ( \overline{p} ) } (t), | n ^{\prime} \rangle \langle n ^{\prime } | \}
+ \sum _{ ( n ^{\prime} n ) } \omega _{ n^{\prime} n } | n \rangle \langle n ^{\prime} | \rho ^{ ( \overline{p} ) } (t) | n ^{ \prime } \rangle \langle n | .
\end{eqnarray}    
By comparing the above equation to Eq. (6), we can see that Eq. (16) is also reduced to the Markoff master equation of Lindblad form in the low-density limit with respect to holes. 

In Ref. \cite{Huang}, Eq. (16) is derived from Eq. (18) after considering the conservation of number of particles in each transition. If we set $A _{ \overline{p} } (t) =0$, we can reduce Eq. (18) to 
\begin{eqnarray}
\frac{\partial}{\partial t} \rho (t) = i [ \rho (t) , H (t) ] + \{ \rho (t) , A _{p} (t) \}. 
\end{eqnarray}
On the other hand, by rewriting Eq. (18) with respect to $\rho ^{ ( \overline{p} ) } (t)$, we have  
\begin{eqnarray}
\frac{\partial}{\partial t} \rho ^{ ( \overline{p} ) }(t) = i [ \rho ^{ ( \overline{p} ) } (t) , H (t) ] + \{ \rho ^{ ( \overline{p} ) } (t) , A _{ \overline{p} } (t) \} 
\end{eqnarray} 
if $A _{p} (t) =0$. The above two equations, in fact, can be obtained by considering the nonhermitian Hamiltonians with $A _{p} (t)$ and $A _{ \overline{p} } (t) $ as imaginary parts for particles and holes, respectively. \cite{Huang} The nonhermitian Hamiltonians have been considered in the quasiparticle theory, in which $A _ {p} (t) $ ($A _{ \overline{p} } (t)$) can correspond to the lifetime of excited particles (holes) above (below) the Fermi energy. \cite{Huang,Kevan,Campillo,Hedin} The loss and gain factors $L _{f} (\rho (t))$ and $G _{f} (\rho (t))$ can be obtained from the last two terms of Eq. (18) by setting $A _{p} (t)= - A _{1} (\rho (t))$ and $A _{ \overline{p} } (t) =- A _{2} (\rho (t))$, so both the lifetimes of particles and holes are incorporated in Eq. (16). \cite{Huang} The master equation, however, may be used to model the time-evolution of the nonequilibrium system with no well-defined (quasi-)Fermi energy. Hence the particles and holes are the occupied and empty parts of any orbitals in the master equation while they describe the filled and empty orbitals above and below Fermi energy in the conventional quasiparticle theory.               

The following equation 
\begin{eqnarray}
\frac{\partial}{\partial t} \rho (t) = i [ \rho (t), H(t) ] - \frac{1}{2} \sum _{l} \{ \rho (t), {\cal V} _{l} [ I - \rho (t) ] {\cal V}_{l} ^{\dagger} \} + \frac{1}{2} \sum _{l} \{ I - \rho (t), {\cal V} _{l} ^{\dagger} \rho (t) {\cal V} _{l} \} 
\end{eqnarray}
can be obtained from Eq. (18) by setting $A _{p} (t) =  - \frac{1}{2} \sum _{l} {\cal V} _{l} [ I - \rho (t) ] {\cal V} _{l} ^{\dagger}$  and $A _{ \overline{p} } (t) = - \frac{1}{2} \sum _{l} {\cal V} _{l} ^{\dagger} \rho (t) {\cal V} _{l}$. Here $\{ {\cal V} _{l} \}$ is a set of operators. \cite{Huang} In the low-density limit with respect to particles (or holes), we can reduce the above equation to the master equation of Lindblad form, of which the relaxation term is given by Eq. (7). Eq. (18), in fact, is a general equation for fermions and can be used to model different Fermi systems. For an example, the density matrix of the form
\begin{eqnarray}
\rho _{q} (t) = \left[
\begin{array}{c}
\rho _{s} (t) \text{ \ \ \ \ \ \  }  \kappa _{s} (t) \\
\text{ \ } - \kappa ^{\ast} _{s} (t) \text{ \ \ \ } I _{s} - \rho _{s} ^{ \ast } (t)
\end{array}
\right]
\end{eqnarray}    
has been introduced in the coordinate space for fermionic quasiparticles in the relativistic Hartree-Bogoliubov model. \cite{Paar,Valatin} Here $\rho _{s} (t)$ is the one-particle density matrix, $\kappa _{s} (t)$ is the antisymmetric pairing tensor, and $I _{s}$ is the one-body identity operator. Such a model is a BCS pairing model for meson-nucleon couplings. The matrix $I _{s} - \rho ^{*} _{s} (t)$ is the conjugate of the matrix $I _{s} - \rho _{s} (t)$ for holes, and $\rho _{q} (t)$ is symmetric with respect to particles and holes. To preserve the symmetric form of $\rho _{q}(t)$ when we model it by Eq. (18), we just need to require \cite{Huang2}
\begin{eqnarray}
SA _{p} ^{\ast}(t)S= A_{\overline{p}} (t) \text{ \ and \ } S =  \left[
\begin{array}{c}
0 \text{ \ \ \ } I _{s} \\
I _{s} \text{ \ \ } 0
\end{array}
\right] 
\end{eqnarray}  
as a constraint on the relaxation term. (In the relativistiv Hartree-Bogoliubov model, the effective Hamiltonian $H(t)$ follows $S H ^{*} (t) S = - H (t)$.) In Ref. \cite{Huang2}, a similar constraint on Eq. (18) is taken into account to extend the time-dependent Bogoliubov equation \cite{superconductivity} for the quasiparitcles in the conventional superconductors. Such an constaint is important to obtain the corresponding semiclassical equation \cite{Entin} in the incoherent limit.        

In the relativisitic Hartree-Bogoliubov model, the density matrix $\rho _{q} (t)$ represents the quasiparticles of which the orbitals are composed of electron and hole parts. The quasiparticles are fermions, so there are corresponding quasiholes. If quasiparticles and quasiholes can couple to form $\lq\lq$new" particles just as how electrons and holes do, it seems natural to introduce the density matrix 
\begin{eqnarray}
\rho _{ q ^{\prime} } (t) = \left[
\begin{array}{c}
\rho _{q} (t) \text{ \ \ \ \ \ \  }  \kappa  _{q}(t) \\
\text{ \ } - \kappa _{q} ^{ \ast } (t) \text{ \ \ \ } I _{q} - \rho _{q} ^{\ast} (t) 
\end{array}
\right]
\end{eqnarray}
for the $\lq \lq$new" particles. Here $I _{q}$ and $\kappa _{q} (t)$ are the identity operator and re-pairing tensor for the quasiparticle (quasihole) orbitals, respectively. To construct the master equation for $\rho _{ q ^{\prime} } (t)$, we shall consider an additional constraint similar to Eq. (61). \cite{Huang2} We note that multiple order parameters \cite{superconductivity2,heavy_fermion} can be incorporated after introducing the re-pairing tensor $\kappa  _{q}(t)$. In addition, both the particle-particle (hole-hole) and particle-hole pairings are taken into account in $\rho _{ q ^{\prime} } (t)$. \cite{Huang2} Two type pairings have been considered to unify BCS theory and antiferromagnetic/ferromagnetic theory. \cite{Lee,Laughlin} Actually a density-matrix series $\{ \rho _{q,j} (t) \}$ can be constructed by \cite{Huang2}
\begin{eqnarray}
\rho _{ q,j+1 } (t) = \left[
\begin{array}{c}
\rho _{q,j} (t) \text{ \ \ \ \ \ \  }  \kappa _{q,j}(t) \\
\text{ \ } - \kappa _{q,j} ^{ \ast } (t) \text{ \ \ \ } I _{q,j} - \rho _{q,j} ^{\ast} (t) 
\end{array}
\right] 
\end{eqnarray}
with $j=1,2,3,...$ as positive integers. Then we can re-obtain Eqs. (60) and (62) by setting $\rho _{q,1} (t) = \rho _{s} (t)$, $\rho _{q,2} (t) = \rho _{q} (t)$, and $\rho _{q,3}(t) = \rho _{ q ^{\prime} } (t)$. The corresponding Hamiltonians form a chain of Hamiltonians \cite{fractal}.  

For noninteracting systems composed of finite identical fermions following the modern quantum mechanics, the trace $tr \rho  (t) $ of the density matrix equals the number of particles and thus $0 < tr \rho (t) < \infty$. It may be reasonable to assume that $H(t)= H_{0}$ as $t \leq t _{0}$, and the system is in an stationary state at $t _{0}$. In such a case, we shall set   
\begin{eqnarray}
\rho (t _{0}) = \sum _{n} c _{n} | n \rangle \langle n | \text{ with } 0 \leq c _{n} \leq 1 \text{ for all } n
\end{eqnarray}
such that the initial density matrix is incoherent with respect to the eigenorbitals of $H_{0}$. Here we require that $tr \rho ( t_{0} ) = \sum _{n} c_{n} $ equals the number of particles. Each $c _{n}$ is the initial occupation number at orbital $| n \rangle$. The validity of $tr \rho (t) = tr \rho (t _{0})$ is expected under Eq. (16) because the conservation of particles is taken into account. \cite{Burke,Huang,Ralph} To prove the invariance of $tr \rho (t)$, we note that the solution to Eq. (16) is fixed under the contraction $F$ introduced in definition 3.4. For any $\Omega (t) \in {\cal S} _{2}$ with $\sup _{t} tr \Omega (t) < \infty$, it is shown in Appendix D that 
\begin{eqnarray}
tr \Lambda (t) = tr \rho ( t _{0} ) + \sum _{ ( n ^{\prime} n ) } \omega _{ n n^{\prime} } \int _{t _{0}} ^{t} dt ^{\prime} \langle n ^{\prime} | \Omega (t ^{\prime}) |n ^{\prime} \rangle ( 1 - \langle n | \Lambda ( t ^{\prime} ) | n \rangle )  
\end{eqnarray}
\[
-\sum _{ ( n ^{\prime} n ) } \omega  _{ n n^{\prime} } \int _{t _{0}} ^{t} dt ^{\prime} ( 1 - \langle n | \Omega ( t ^{\prime} ) | n \rangle ) \langle n ^{\prime} | \Lambda (t ^{\prime}) |n ^{\prime} \rangle
\]
if $\Lambda (t) = F(\Omega (t))$ when $\rho ( t _{0} )$ following Eq. (64) is of a finite trace. If $tr \rho (t)$ is bounded, we can set $\Omega (t) = \rho (t) = F(\rho (t) ) = \Lambda (t)$ in the above equation to prove $tr \rho (t) = tr \rho (t _{0})$. To see that $ tr \rho (t) $ is finite, we can ignore the third term of the above equation to obtain the   inequality  
\begin{eqnarray}
0 < tr \Lambda (t) < tr \rho ( t _{0} ) + \sum _{ ( n ^{\prime} n ) } \omega _{ n n^{\prime} } \int _{t _{0}} ^{t} dt ^{\prime} \langle n ^{\prime} | \Omega (t ^{\prime}) |n ^{\prime} \rangle < tr \rho ( t _{0} ) + \frac{1}{2} \sup _{t} tr \Omega (t) 
\end{eqnarray} 
since every operator in ${\cal S} _{2}$ is positive and we choose $t_{f}< t_{0} + 1/4M$. Based on the above equation, we can prove that the trace of each $ \rho _{j} (t) $ constructed by Eqs. (53) and (54) is smaller than $2 tr \rho ( t _{0} ) $ by induction. Therefore, we just need to note $0< tr \rho (t) \leq \underline{lim} _{j} tr \rho _{j} (t) < 2 tr \rho ( t _{0} )$ to complete the proof.                       

\section{Conclusion}

In conclusion, the positivity and Pauli's exclusion principle are both preserved under the nonlinear quantum master equation introduced in Refs. \cite{Huang} and \cite{Ralph} when there exists an upper bound for the transition rate. Both the loss and gain factors of the equation induce the decoherence, which is important to the positivity and Pauli's exclusion principle. The number of particles is conserved if the initial density matrix is of a finite trace. Such an equation can be generalized to model BCS-type quasiparticles. On the other hand, it can be reduced to a Markoff master equation of Lindblad form in the low-density limit with respect to particles or holes.     
 
\section*{Appendix A}

For any two kets $| \alpha \rangle$ and $| \beta \rangle$ and a time-dependent operator ${\cal Q} (t) \in {\cal S} _{0}$, both $Re \langle \alpha | {\cal Q} ( t ^{\prime} ) | \beta \rangle$ and $Im \langle \alpha | {\cal Q} ( t ^{\prime} ) | \beta \rangle$ are Riemann integrable on $[t_{0},t_{f}]$ because they are continuous (real) functions. Here $Re \langle \alpha | {\cal Q} ( t ^{\prime} ) | \beta \rangle$ and $Im \langle \alpha | {\cal Q} ( t ^{\prime} ) | \beta \rangle$ denote the real and imaginary parts of $\langle \alpha | {\cal Q} ( t ^{\prime} ) | \beta \rangle$. Hence $\int _{ t_{1} } ^{ t_{2} } d t ^{\prime} \langle \alpha | {\cal Q} ( t ^{\prime} ) | \beta \rangle = \int _{ t_{1} } ^{ t_{2} } d t ^{\prime} Re \langle \alpha | {\cal Q} ( t ^{\prime} ) | \beta \rangle + i \int _{ t_{1} } ^{ t_{2} } d t ^{\prime} Im \langle \alpha | {\cal Q} ( t ^{\prime} ) | \beta \rangle$ is well-defined for any $t_{1}$ and $t_{2} \in [t_{0}, t_{f}]$. In addition, the operator $ {\cal O} = \int _{ t_{1} } ^{  t_{2} } dt ^{\prime}{\cal Q} ( t ^{\prime} )$ is well-defined in the sense that $\langle \alpha |{\cal O}| \beta \rangle = \int _{ t_{1} } ^{  t_{2} } dt ^{\prime}\langle \alpha |{\cal Q} ( t ^{\prime} )| \beta \rangle$ for any $| \alpha \rangle$ and $| \beta \rangle$. The time-derivative on each operator $\Omega ^{\prime} (t) \in {\cal S} ^{\prime} _{1}$ is defined by Eqs. (30) and (31) after introducing the intergal on ${\cal S} _{0}$. Such a definition, however, cannot be extended to ${\cal S} _{1}$ when the unitary operator $U(t)$ is generated by an unbounded Hamiltonian $H(t)$.           

The following corollaries can be used to simply some calculations in section III. \newline
{\bf Corollary A1} $Let$ ${\cal Q} _{3} (t) = {\cal Q} _{1} (t) {\cal Q} _{2} (t)$ $for$ $any$ $two$ ${\cal Q} _{1} (t)$ $and$ ${\cal Q} _{2} (t) \in {\cal S} _{0}$. $Then$ ${\cal Q} _{3} (t) \in {\cal S} _{0}$. ($Similarly$, $if$ $\Omega ^{\prime} _{3} (t) = \Omega ^{\prime} _{1} (t) \Omega ^{\prime} _{2} (t)$ $for$ $any$ $two$ $\Omega ^{\prime} _{1} (t)$ $and$ $\Omega ^{\prime} _{2} (t) \in {\cal S} ^{\prime} _{1}$, $we$ $have$ $\Omega ^{\prime} _{3} (t) \in {\cal S} ^{\prime} _{1}$). \newline
{\bf proof} Since $|||{\cal Q} _{3} (t)||| \leq |||{\cal Q} _{1}(t) ||| \times ||| {\cal Q} _{2} (t)|||$, we just need to show that $lim _{ t_{2} \rightarrow t _{1} } \Vert [ {\cal Q} _{3} ( t _{2} ) - {\cal Q} _{3} ( t _{1} ) ] | \alpha \rangle \Vert =0 $ for all $| \alpha \rangle $ at any time $t _{1} \in [t _{0}, t_{f}]$. By setting $| \beta \rangle = {\cal Q}_{2} (t _{1}) | \alpha \rangle$, we have $lim _{ t_{2} \rightarrow t _{1} } \Vert [ {\cal Q} _{3} ( t _{2} ) - {\cal Q} _{3} ( t _{1} ) ] | \alpha \rangle \Vert \leq lim _{ t_{2} \rightarrow t _{1} } \Vert [ {\cal Q} _{1} ( t _{2} ) - {\cal Q} _{1} ( t _{1} ) ] | \beta \rangle \Vert + lim _{ t_{2} \rightarrow t _{1} } \Vert  {\cal Q} _{1} ( t _{2} ) [ {\cal Q} _{2} ( t _{2} ) - {\cal Q} _{2} ( t_{1} ) ] | \alpha \rangle \Vert = 0$. \text{ \ \ \ \ \ \ \ \ \ \ \ \ \ \ \ \ \ \ \ \ \ \ \ \ \ \ \ \ \ \ \ \ \ \ \ \ \ \ \ \ \  \ \ \ \ \ \ \ \ \ \ \ \ }  {\bf QED} 
\newline \newline
{\bf Corollary A2} $\text{ \ }$ $Let$ $\Omega ^{\prime} _{1} (t)$ $and$ $\Omega ^{\prime} _{1} (t)$ $be$ $two$ $operators$ $in$ ${\cal S} ^{\prime} _{1}$, $and$ $assume$ $that$ $we$ $can$ $find$ $two$ $operators$ ${\cal Q} _{1} (t)$ $and$ ${\cal Q} _{2} (t)$ in ${\cal S} _{0}$ $such$ $that$ ${\cal Q} _{1} (t) = \frac{\partial}{\partial t} \Omega ^{\prime} _{1} (t)$ $and$ ${\cal Q} _{2} (t) = \frac{\partial}{\partial t} \Omega ^{\prime} _{2} (t)$ $on$ $[t_{1},t_{2}] \subseteq [t _{0} , t _{f}]$. $Then$ $\Omega ^{\prime} _{3} (t) \equiv \Omega ^{\prime} _{1} (t) \Omega ^{\prime} _{2} (t) \in {\cal S} _{1} ^{\prime}$ $follows$ $ \frac{\partial}{\partial t} \Omega ^{\prime} _{3} (t) = {\cal Q} _{1} (t) \Omega ^{\prime} _{2} (t) + \Omega ^{\prime} _{1} (t) {\cal Q} _{2} (t) $ $when$ $t \in [ t _{1}, t _{2} ]$. $That$ $is$, $\frac{\partial}{\partial t} (\Omega ^{\prime} _{1} (t) \Omega ^{\prime} _{2} (t)) = ( \frac{\partial}{\partial t} \Omega ^{\prime} _{1} (t)) \Omega ^{\prime} _{2} (t) + \Omega ^{\prime} _{1} (t) (\frac{\partial}{\partial t}\Omega ^{\prime} _{2} (t)$).  \newline
{\bf proof} $\text{ \ }$ For any $|\alpha\rangle$, we can define two-parameter kets $ | \alpha _{1} ( t , t ^{\prime} ) \rangle \equiv {\cal Q} _{1} ( t ) \Omega _{2} ^{\prime} ( t ^{\prime} ) | \alpha \rangle $ and $| \alpha _{2} ( t , t ^{\prime} ) \rangle \equiv \Omega _{1} ^{\prime} ( t ) {\cal Q} _{2} ( t ^{\prime} )  | \alpha \rangle $ as $(t,t ^{\prime}) \in X = [t_{1}, t_{2}] \times [ t_{1}, t_{2}]$, which is compact under Euclidean metric. Both $| \alpha _{1} ( t , t ^{\prime} ) \rangle $ and $| \alpha _{2} ( t , t ^{\prime} ) \rangle$ are (uniformly) contiuous such that $ lim _{\Delta t \rightarrow 0, \Delta t ^{\prime} \rightarrow 0} \Vert | \alpha _{1} ( t + \Delta t , t ^{\prime} + \Delta t ^{\prime} ) \rangle - | \alpha _{1} ( t , t ^{\prime} ) \rangle \Vert = lim _{\Delta t \rightarrow 0, \Delta t ^{\prime} \rightarrow 0} \Vert  | \alpha _{2} ( t + \Delta t , t ^{\prime} + \Delta t ^{\prime} ) \rangle  - \alpha _{2} ( t , t ^{\prime} ) \rangle \Vert =0$ on the compact domain $X$. Then for any ket $| \beta \rangle$, 
\[
\text{ \ \ \ \ \ \ \ \ \ \ \ \ \ } \frac{1}{\Delta t} \langle \beta |( \Omega ^{\prime} _{3} (t+\Delta t) -  \Omega ^{\prime} _{3} (t) ) | \alpha \rangle \text{ \ \ \ \ \ \ \ \ \ \ \ \ \ \ \ \ \ \ \ \ \ \ \ \ \ \ \ \ \ \ \ \ \ \ \ \ \ \ \ \ \ \ \ \ \ \ \ \ \ \ \ \ \ \ \ \ \ \ (A1) }      
\]
\[
= \frac{1}{\Delta t}\langle \beta | \Omega ^{\prime} _{1} (t+\Delta t) ( \Omega ^{\prime}_{2} (t+\Delta t) -\Omega ^{\prime} _{2} (t) ) + ( \Omega ^{\prime}_{1} (t+\Delta t) - \Omega ^{\prime} _{1} (t) ) \Omega ^{\prime} _{2} (t) ]| \alpha \rangle \text{ \ }
\]
\[
=\frac{1}{\Delta t} \langle \beta | \Omega ^{\prime} _{1} (t+\Delta t)( \int _{t} ^{ t + \Delta t } dt ^{\prime}  {\cal Q} _{2} ( t ^{\prime} ) ) | \alpha \rangle + \frac{1}{\Delta t} \langle \beta | ( \int _{t} ^{ t + \Delta t } d t ^{\prime} {\cal Q} _{1} ( t ^{\prime} ) ) \Omega ^{\prime} _{2} (t) | \alpha \rangle 
\]
\[
=\frac{1}{\Delta t} \int _{t} ^{ t + \Delta t } dt ^{\prime} ( \langle \beta | \alpha _{2} (t + \Delta t, t ^{\prime}) \rangle + \langle \beta | \alpha _{1} (t ^{\prime} , t) \rangle ) \text{ \ \ \ \ \ \ \ \ \ \ \ \ \ \ \ \ \ \ \ \ \ \ \ \ \ \ \ \ \ \ \ \ \ \ } 
\]
\[
\rightarrow  \langle \beta | \alpha _{2} (t, t ) \rangle + \langle \beta | \alpha _{1} (t , t) \rangle \text{ as } \Delta t \rightarrow 0. \text{ \ \ \ \ \ \ \ \ \ \ \ \ \ \ \ \ \ \ \ \ \ \ \ \ \ \ \ \ \ \ \ \ \ \ \ \ \ \ \ \ \ \ \ \ \ \ }
\] 
Hence 
\[
\text{  \ \ \ \ \ \ \ \ \ \ \ \ \ \ \ \  \ \ \ \ \ \ \ \ \ } \frac{d}{d t} \langle \beta | \Omega _{3} (t) | \alpha \rangle = \langle \beta | [ {\cal Q} _{1} (t) \Omega ^{\prime} _{2} (t) + \Omega ^{\prime} _{1} (t) {\cal Q} _{2} (t) ] | \alpha \rangle  \text{ \ \ \ \ \ \ \ \ \ \ \ \ \ \ \ \ \ \ \ \ \ \ (A2) }
\]
for any $ | \alpha \rangle \text{ and } | \beta \rangle$ when $t \in [t _{1}, t _{2} ] $, and we can complete the proof because we have $\Omega _{3} (t) \in {\cal S} ^{\prime} _{1}$ and ${\cal Q} _{1} (t) \Omega ^{\prime} _{2} (t) + \Omega ^{\prime} _{1} (t) {\cal Q} _{2} (t) \in {\cal S} _{0}$ from corrollary A1. $\text{\ \ \ \ \ \ \ \ \ \ \ \ \ \ \ \ \ \ \ \ \ \ \ \ \ }$ {\bf QED} \newline \newline
{\bf Corollary A3} $Let$ ${\cal S} _{0} ^{(2)}$ $be$ $the$ $vector$ $space$ $composed$ $of$ $all$ $the$ $mappings$ ${\cal Q} ( t ; t ^{\prime} )$ $from$ $ (t , t ^{\prime}) \in [t _{0}, t_{f} ] \times [ t _{0}, t_{f} ]$ $to$ $bounded$ ($linear$) $operators$ $such$ $that$ $\sup \Vert {\cal Q} (t ; t ^{\prime})\Vert < \infty $ and $lim _{ \Delta t , \Delta t ^{ \prime } \rightarrow 0} \Vert [ {\cal Q} ( t + \Delta t ; t^{ \prime } + \Delta t ^{ \prime } ) -{\cal Q} (t;t^{\prime}) ]| \alpha \rangle \Vert =0$ $for$ $all$ $| \alpha \rangle$. $Then$ $each$ ${\cal Q} (t;t ^{\prime}) \in {\cal S} _{0} ^{(2)}$ $is$ $intergrable$ $with$ $respect$ $to$ $t$ $and/or$ $t ^{\prime}$. $In$ $addition$, ${\cal Q} _{3} (t;t^{\prime}) = {\cal Q} _{1} ( t ; t ^{\prime} ){\cal Q} _{2} ( t ; t ^{\prime} ) \in {\cal S} _{0} ^{(2)} $ if both ${\cal Q} _{1} ( t ; t ^{\prime} )$ and ${\cal Q} _{2} ( t ; t ^{\prime} ) \in {\cal S} _{0} ^{ (2) }$. \newline \newline
{\bf Corollary A4} $\text{ \ }$ $Let$ ${\cal Q}(t;t^{\prime})$ $be$ $a$ $two$-$parameter$ $operator$ $in$ ${\cal S} _{0} ^{(2)}$, $which$ $is$ $introduced$ $in$ $corollary$ $A3$. $Assume$ $that$ ${\cal O}( t ; t^{\prime} )$ $follows$ ${\cal O}( t ; t^{\prime} ) = {\cal O} _{0} ( t ^{\prime} ) + \int _{ t ^{\prime} } ^{t} dt ^{ \prime \prime } {\cal Q} ( t ^{ \prime \prime } ; t ^{\prime} )$ $at$ $t \geq t ^{\prime}$ and ${\cal O}( t ; t^{\prime} ) = {\cal O} _{0} ( t ^{\prime} )$ $at$ $t \leq t ^{\prime}$ $for$ $some$ ${\cal O} _{0} (t) \in {\cal S} _{0}$. $Then$ $the$ $operator$ $W (t) \equiv \int _{ t_{0} } ^{t} d t ^{\prime} {\cal O}(t;t^{\prime}) \in {\cal S} ^{\prime} _{1}$,  $and$ $\frac{\partial}{\partial t} W (t) = {\cal O} _{0} (t) + \int _{ t _{0} } ^{t} dt ^{\prime} {\cal Q} (t,t ^{\prime}) \in {\cal S} _{0}$. \newline
{\bf proof} $\text{ \ }$ It is easy to prove that $W(t)$ is well-defined because ${\cal O} (t _{1} ; t ) \in {\cal S} _{0}$ for each $t _{1}$. For any two kets $| \alpha \rangle$ and $| \beta \rangle$, 
\[
\text{ \ \ \ \ \ \ \ \ \ \ \ \ \ \ \ \ \ \ } \langle \beta | W (t) | \alpha \rangle = \int _{ t_{0} } ^{t} dt ^{\prime}  \langle \beta | {\cal O} ( t ; t ^{\prime}) | \alpha \rangle \text{ \ \ \ \ \ \ \ \ \ \ \ \ \ \ \ \ \ \ \ \ \ \ \ \ \ \ \ \ \ \ \ \ \ \ \ \ \ \ \ \ \ \ \ \ \ \ \ \ \ (A3) }
\]
\[
\text{ \ \ \ \ \ \ \ \ \ \ \ \ \ \ \ \ \ } = \int _{ t _{0} } ^{t} dt ^{\prime } \langle \beta | {\cal O} _{0} (t ^{\prime}) | \alpha \rangle + \int _{ t _{0} \leq t ^{\prime} \leq t ^{ \prime \prime } \leq t } dt ^{\prime } dt ^{ \prime \prime } \langle \beta | {\cal Q} (t ^{\prime  \prime} ; t ^{\prime}) | \alpha \rangle 
\]
\[
\text{ \ \ \ } =\langle \beta | \{ \int _{ t_{0} } ^{t} d t^{\prime} [{\cal O} _{0} ( t ^{\prime} ) + \int _{ t_{0} } ^{ t ^{\prime} } d t ^{ \prime \prime} {\cal Q} ( t ^{ \prime } ; t ^{ \prime \prime } ) ] \} | \alpha \rangle. 
\]
Then we have $\frac{\partial}{\partial t} W (t) = {\cal O} _{0} (t) + \int _{ t _{0} } ^{t} dt ^{\prime} {\cal Q} (t;t ^{\prime}) \in {\cal S} _{0}$ because $|| \int _{ t_{0} } ^{t} d t ^{\prime} {\cal Q}(t;t^{\prime}) || \leq |t_{f} - t_{0}| \times \sup ||{\cal Q} ( t ; t ^{\prime} )|| $ and $lim _{\Delta t \rightarrow 0} \Vert ( [ \int _{ t_{0} } ^{ t + \Delta t } d t ^{\prime} {\cal Q}( t + \Delta t ; t ^{\prime} ) - \int _{ t_{0} } ^{t} d t ^{\prime} {\cal Q}(t;t^{\prime})] | \alpha\rangle \Vert =0$ for any $| \alpha \rangle $. Since $ \Vert W ( t _{2} ) -  W ( t _{1} ) \Vert \leq |t _{2} -t _{1}| \times |||\frac{\partial}{\partial t} W (t) |||$, we have $W (t) \in {\cal S} ^{\prime} _{1}$.  $\text{ \ }$ {\bf QED} \newline \newline
{\bf Corollary A5} $Let$ $\Omega ^{\prime} (t) \in {\cal S} ^{\prime} _{1}$ $and$ ${\cal Q}(t) \in {\cal S} _{0}$ $be$ $two$ $time$-$dependent$ $operators$ $following$ $\frac{\partial}{\partial t} \Omega ^{\prime} (t) = {\cal Q} (t)$ $on$ $the$ $interval$ $[ t_{1} , t_{2}] \subseteq [t_{0}, t_{f}]$. $If$ ${\cal Q} ^{\dagger}(t) \in {\cal S} _{0}$, $the$ $equation$ $\frac{\partial}{\partial t} \Omega ^{\prime \dagger} (t) = {\cal Q} ^{\dagger} (t)$ $is$ $valid$ on $[ t_{1} , t_{2}]$. \newline 

\section*{Appendix B} 
To prove the existence and uiqueness of $K ^{\prime} (t;t ^{\prime})$ satisfying Eq. (35), we can construct a family of mappings $\{ {\cal G} _{ t ^{\prime} } \}$ from ${\cal S} ^{\prime} _{1}$ to ${\cal S} ^{\prime} _{1}$ itself such that 
\[
\text{ \ \ \ \ \ \ \ \ \ \ \ \ \ \ \ \ \ \ \ \ \ \ \ \ \ \ \ \ \ \ } \Lambda ^{\prime} (t) = I - \int _{ t ^{\prime} } ^{t} dt ^{ \prime \prime } {\cal A} _{U} ( t ^{ \prime \prime } ) \Omega ^{\prime} (t ^{ \prime \prime }) \text{ \ \ \ \ \ \ \ \ \ \ \ \ \ \ \ \ \ \ \ \ \ \ \ \ \ \ \ \ \ \ \ (B1) }
\]
at $ t \geq t ^{\prime} $ and $\Lambda ^{\prime} (t) = I $ at $ t \leq t ^{\prime} $ iff $\Lambda ^{\prime} (t) = {\cal G} _{t ^{\prime} } ( \Omega ^{\prime} (t))$. Such a family of mappings are parametrized by $t ^{\prime} \in [t_{0},t_{f}] $. Each $ {\cal G} _{ t ^{\prime} }$ is well-defined because the integrand $ - {\cal A} _{U} ( t ) \Omega ^{\prime} (t) \in {\cal S} _{0}$. Since
\[
\text{ \ \ \ \ \ \ \ \ \ \ \ \ \ \ \ \ \ \ \ } ||| {\cal G} _{ t ^{ \prime} } ( \Omega ^{\prime} _{1} (t) ) - {\cal G} _{ t ^{ \prime} } ( \Omega ^{\prime} _{2} (t) ) |||
< \frac{1}{2} ||| \Omega ^{\prime} _{1} (t) - \Omega ^{\prime} _{2} (t) |||, \text{ \ \ \ \ \ \ \ \ \ \ \ \ \ \ \ \ \ \ (B2) }
\]
$\{ {\cal G} _{ t^{\prime} } \}$ is a family of contractions on ${\cal S} ^{\prime} _{1}$. From the fixed point theory, there exists a unique $K ^{\prime} (t;t^{\prime}) \in {\cal S} ^{\prime} _{1}$ such that $K ^{\prime} ( t ; t ^{\prime} )= {\cal G} _{ t ^{\prime} } ( K ^{\prime} (t;t^{\prime}) )$ at each $t ^{\prime}$. In addition, we can introduce $K (t;t^{\prime}) $ just as in lemma 3.2. If the Hamiltonian $H(t)$ is bounded, $||| \frac{\partial}{\partial t} K (t;t^{\prime}) ||| $ is bounded and Eq. (36) can be defined by Eqs. (30) and (31). Although Eq. (36) suffers unbounded problem when $H(t)$ is not bounded, we can define $K (t;t^{\prime})$ by Eq. (35) after tranforming $K (t;t^{\prime})$ to $K ^{\prime} (t; t^{\prime})$. 

Direct estimations yield  
\[  
\text{ \ \ \ \ \ \ \ \ \ \ \ \ \ \ \ \ \ \ \ \ \ \ \ \ \ \ \ \ \ \ \ } |||{\cal G} _{ t ^{\prime} } ( \Omega ^{\prime} (t) ) ||| < 1 + \frac{1}{2} |||\Omega ^{\prime} (t)||| \text{ \ \ \ \ \ \ \ \ \ \ \ \ \ \ \ \ \ \ \ \ \ \ \ \ \ \ \ \ \ \ \ (B3) }   
\]
\[
\text{ \ \ \ \ \ \ \ \ \ \ \ \ \ \ \ \ \ } |||{\cal G} _{ t ^{\prime} _{2} } ( \Omega ^{\prime} (t) ) - {\cal G} _{ t _{1} ^{\prime} } ( \Omega ^{\prime} (t) )||| \leq  2M |t_{2} ^{\prime} - t _{1} ^{\prime} | \times ||| \Omega ^{\prime} (t) |||. \text{ \ \ \ \ \ \ \ \ \ \ \ \ \ \ \ \ (B4) }   
\] 
From Eq. (B3), we can prove that 
\[
\text{ \ \ \ \ \ \ \ \ \ \ \ \ \ \ \ \ \ \ \ \ \ \ \ \ \ \ \ \ \ \ \ \ \ \ \ \ } \sup _{ t,t^{\prime} } \Vert K ^{\prime} (t;t ^{\prime})\Vert \leq 2. \text{  \ \ \ \ \ \ \ \ \ \ \ \ \ \ \ \ \ \ \ \ \ \ \ \ \ \ \ \ \ \ \ \ \ \ \ \ \ \ \ (B5)}
\] 
(Actually $\sup _{ t,t^{\prime} } \Vert K ^{\prime} (t;t ^{\prime})\Vert =1$ because ${\cal A} _{U} (t)$ is positive.) In addition, we have
\[
\text{ \ \ \ \ \ \ \ \ \ \ \ \ \ \ \ \ \ \ \ \ \ \ \ \ \ } \sup _{t} \Vert  K ^{\prime} (t;t ^{\prime} _{2} ) -  K ^{\prime} (t;t ^{\prime} _{1}) \Vert \leq 8M |t_{2} ^{\prime} - t _{1} ^{\prime}| \text{ \ \ \ \ \ \ \ \ \ \ \ \ \ \ \ \ \ \ \ \ \ \ \ \ (B6) } 
\]
from Eqs. (B2) and (B4) because $K ^{\prime} (t;t ^{\prime})$ is fixed by ${\cal G} _{ t ^{\prime} }$. Eq. (B6) indicates that $K ^{\prime} (t;t^{\prime})$ can be taken as as a mapping from $ t \in [ t_{0} , t _{f} ] $ to ${\cal S} ^{\prime} _{1}\subseteq {\cal S} _{0}$, so $K ^{\prime} (t;t^{\prime})$ is integrable with respect to $t$ and/or $t ^{\prime}$. 

Since $K ^{\prime} (t;t^{\prime}) \in {\cal S} ^{\prime} _{1}$ at each $t ^{\prime} \in [t_{0},t_{f}]$, we have $lim _{\Delta t, \Delta t^{\prime} \rightarrow 0 } \Vert [ K ^{\prime} (t+ \Delta t;t ^{\prime} + \Delta t ^{\prime} ) - K ^{\prime} (t;t^{\prime}) ] \Vert \leq lim _{\Delta t,\Delta t^{\prime} \rightarrow 0 } \Vert [ K ^{\prime} (t+\Delta t;t ^{\prime} + \Delta t ^{\prime} ) - K ^{\prime} (t+\Delta t ; t^{\prime}) ] \Vert + lim _{\Delta t,\Delta t^{\prime} \rightarrow 0 } \Vert [ K ^{\prime} (t+\Delta t;t ^{\prime} ) - K ^{\prime} (t;t^{\prime}) ] \Vert =0$ at any $(t,t^{\prime})$ in the definition domain from Eq. (B6). Together with Eq. (B5), therefore, we can prove that $K ^{\prime} (t;t^{\prime}) \text{ and } K ^{\prime \dagger} (t;t^{\prime}) \in {\cal S} _{0} ^{ (2) }$, which is introduced in corollary A3. It is easy to see that $K (t;t^{\prime}) \text{ and } K ^{\dagger} (t;t^{\prime})$ are also in ${\cal S} _{0} ^{ (2) }$ from the definition of $K (t;t^{\prime})$. From corollary A3, we can prove that the integrals in Eqs. (37) and (38) are well-defined because ${\cal B} _{U} (t^{\prime})$ and ${\cal B} (t ^{\prime})$ can be taken as operators in ${\cal S} ^{(2)} _{0}$.  

\section*{Appendix C}

In this Appendix, we assume that $\Omega (t) \in {\cal S} _{2}$. Since $A _{1} (\Omega (t))$ and $A _{2} (\Omega (t))$ are both diagonalized by the eigenorbitals $\{ | n \rangle \}$ of $H_{0}$, we have $|||A _{1} (\Omega (t))||| = \sup _{n,t} | \langle n | A _{1} (\Omega (t)) | n \rangle | \leq M$ and $||| A _{2} (\Omega (t)) |||= \sup _{n,t} | \langle n | A _{2} (\Omega (t)) | n \rangle | \leq M$ after some calculations. To show that $A _{1} (\Omega (t))$ and $A _{2} (\Omega (t)) \in {\cal S} _{0}$, therefore, we just need to prove that 
\[ 
\text{ \ \ \ \ \ \ \ \ \ \ \ \ \ \ \ \ \ \ \ \ \ \ } lim _{ t_{2} \rightarrow t _{1} } \Vert A _{j} ( \Omega ( t _{2} ) ) | \alpha \rangle - A _{j} ( \Omega ( t _{1} ) ) | \alpha \rangle \Vert =0 \text{ \ \ \ \ \ \ \ \ \ \ \ \ \ \ \ \ \ \ \ \ \ \ \ \ (C1) } 
\]   
for all $| \alpha \rangle$ as $j=1$ and $2$. It is easy to see the validity of the above equation if $| \alpha \rangle $ is in the set composed of all the finite linear combinations of eigenorbitals of $H_{0}$. Because such a set is dense and both $|||A _{1} (t) |||$ and $|||A _{2} (t) |||$ are bounded by $M$, we can prove that Eq. (C1) holds true for any $|\alpha\rangle$.         
  
\section*{Appendix D}

In this Appendix, we assume that $\rho ( t _{0} )$ satisfies Eq. (64). To prove Eq. (65), in which $\Lambda (t) = F(\Omega (t))$, it is important to estimate $tr {\cal K} _{\Omega} (t,t _{a} ) |n \rangle \langle n| {\cal K} _{\Omega} ^{\dagger} (t ; t _{a} ) = \langle n | {\cal K} _{\Omega} ^{\dagger} (t;t _{a} ) {\cal K} _{\Omega} (t;t _{a} ) | n \rangle $ for each normalized eigenorbital $| n \rangle $ of $H(t)$ because both $\rho (t _{0})$ and $A _{2} (\Omega (t))$ in Eq. (45) are diagonalized by $\{ | n \rangle \}$. Here we assume that $ t_{0} \leq t _{a} \leq t$. Because $\langle n | {\cal K} _{\Omega} ^{\dagger} (t;t _{a} ) {\cal K} _{\Omega} (t;t _{a} ) | n \rangle = \langle n | U ^{\dagger} (t _{a}) {\cal K} _{\Omega} ^{\prime \dagger} (t;t _{a} ) {\cal K} ^{\prime} _{\Omega} (t;t _{a} ) U (t _{a}) | n \rangle$, $\frac{\partial}{\partial t} \langle n | {\cal K} _{\Omega} ^{\dagger} (t;t _{a} ) {\cal K} _{\Omega} (t;t _{a} ) | n \rangle$ is well-defined and equals $-2 \langle n | {\cal K} _{\Omega} ^{\dagger} (t;t _{a} ) [ A _{1} ( \Omega (t) ) + A _{2} ( \Omega (t) )]{\cal K} _{\Omega} (t;t _{a} ) | n \rangle $. So  
\[
\text{ \ \ \ \ \ \ \ \ \ } tr {\cal K} _{ \Omega } (t;t _{a} ) | n \rangle \langle n | {\cal K} _{ \Omega } ^{\dagger} (t;t _{a} ) = 1 - \sum _{ (n ^{\prime} n ^{ \prime \prime } ) } \int _{ t _{a} } ^{t} d t ^{\prime} | \langle n ^{\prime}  | {\cal K} _{\Omega} ( t ^{\prime} ; t _{a}  ) |n \rangle | ^{2} \text{ \ \ \ \ \ \ \ \ \ \ \ \ \ (D1) }
\]
\[
\times [ \omega _{ n ^{\prime \prime} n ^{\prime} } (1 - \langle n ^{\prime \prime} | \Omega (t ^{\prime}) | n ^{\prime \prime} \rangle ) + \omega _{ n ^{\prime} n ^{\prime \prime} } \langle n ^{\prime \prime} | \Omega ( t ^{\prime} ) | n ^{\prime \prime} \rangle].            
\]
Let a (time-independent) operator ${\cal C} = \sum _{ n } C_{n} |n \rangle \langle n |$ with $\{ C_{n} \}$ as a set of positive real numbers such that $\sum _{n} C _{n} < \infty$. Such an operator is diagonalized by the eigenorbitals of $H _{0}$. We can extend the above equation as 
\[
\text{ \ \ \ \ } tr {\cal K} _{ \Omega } (t;t _{a} ) {\cal C} {\cal K} _{ \Omega } ^{\dagger} (t;t _{a} ) = tr {\cal C} -  \sum _{ ( n ^{ \prime } n ^{\prime \prime} ) } \int _{ t _{a} } ^{t} d t ^{\prime} \langle n ^{\prime} | {\cal K} _{\Omega} ( t ^{\prime} ; t _{a}  ) {\cal C} {\cal K} _{\Omega} ^{\dagger} ( t ^{\prime} ; t _{a} ) |n ^{\prime}\rangle  \text{ \ \ \ \ \ \ \ \ (D2)} 
\]
\[
\times [ \omega _{ n ^{\prime \prime } n ^{\prime} } (1 - \langle n ^{\prime \prime } | \Omega (t ^{\prime}) | n ^{\prime \prime } \rangle ) + \omega _{ n ^{\prime} n ^{ \prime \prime } } \langle n ^{ \prime \prime } | \Omega ( t ^{\prime} ) | n ^{ \prime \prime } \rangle].            
\]

Let $\{ {\cal C} _{j} \}$ be a set of finite (time-independent) operators parametrized by $j$ such that ${\cal C} _{j} =\sum _{n}  C _{n} ^{(j)} | n \rangle \langle n |$. Here $\{ C _{n} ^{(j)} \}$ is a countable set of positive real numbers such that $\sum _{n} C _{n} ^{ (j) } < \infty$ for each $j$. Let
\[
\text{ \ \ \ \ \ \ \ \ \ \ \ \ \ \ \ \ \ \ \ \ \ \ \ \ \ \ \ \ \ \ \ \ } \Gamma (t)= \sum _{ t _{j} \leq t } tr {\cal K} _{\Omega} (t;t _{\alpha} ) {\cal C} _{j} {\cal K} _{\Omega} ^{\dagger} (t;t _{\alpha} ). \text{  \ \ \ \ \ \ \ \ \ \ \ \ \ \ \ \ \ \ \ \ \ \ \ \ (D3)}
\]
We can prove that  
\[
\text{ \ \ \ \ \ \ } tr \Gamma (t)  = \sum _{ t _{j} \leq t } tr {\cal C} _{j} - \sum _{ ( n ^{\prime} n ) } \int _{ t _{0} } ^{t} d t ^{\prime} \langle n | \Gamma ( t ^{\prime} ) | n \rangle [ \omega _{ n ^{\prime} n } ( 1- \langle n ^{\prime} | \Omega (t ^{\prime})| n ^{\prime} \rangle )    \text{ \ \ \ \ \ \ \ \ \ \ \ (D4) }         
\]
\[
+ \omega _{ n n ^{\prime} } \langle n ^{\prime} | \Omega (t ^{\prime}) | n ^{\prime} \rangle ]
\]
from Eq. (D2). Here each time $t _{j} \geq t _{0}$. Actually Eq. (45) yields a form similar to that provided by Eq. (D3) because $F( \Omega (t) )$ equals ${\cal K} _{\Omega} (t;t _{0} ) \rho (t _{0}) {\cal K} _{\Omega} ^{\dagger} (t;t _{0} )$ plus an integral term, which can be taken as a limit of a summation over ${\cal K} _{\Omega} (t;t ^{\prime} ) A _{2} ( \Omega ( t ^{\prime} ) ) {\cal K} _{\Omega} ^{\dagger} (t;t ^{\prime})$. Because $\sup _{t} tr A _{2} ( \Omega (t) ) \leq M \times \sup _{t} tr \Omega (t)$ for any $\Omega (t) \in {\cal S} _{2}$, we can generalize Eq. (D4) to prove Eq. (65) when both $tr \rho (t _{0})$ and $\sup _{t} tr \Omega (t)$ are finite.

\end{document}